\documentclass[a4paper,USenglish,cleveref,authorcolumns]{lipics-v2021}

\hideLIPIcs

\pdfoutput=1

\usepackage[T1]{fontenc}
\usepackage[utf8]{inputenc}

\usepackage{graphicx,wrapfig}
\usepackage[boxed,ruled,vlined,linesnumbered,noresetcount]{algorithm2e}
\usepackage{microtype}
\usepackage{amsmath}
\usepackage{amsthm}
\usepackage{mathtools}
\usepackage{url}
\usepackage{floatpag}
\usepackage[font=small,labelfont=bf]{caption}

\usepackage{tikz}
\usetikzlibrary{arrows.meta,math,decorations.pathreplacing}

\usepackage{pdfpages}
\usepackage{graphicx}
\usepackage{xspace}
\usepackage{afterpage}
\usepackage{tabularx}
\usepackage[normalem]{ulem}

\newcommand{\etal}{\emph{et al.\@}\xspace}
\newcommand{\ie}{\emph{i.e.,}\xspace}

\newcommand{\Ie}{\emph{I.e.,}\xspace}
\newcommand{\eg}{\emph{e.g.,}\xspace}

\newcommand\bull{{\operatorname{-\xspace}}}
\newcommand{\bigO}[1]{\mathcal{O}\text{(}{#1}\text{)}\xspace}   

\makeatletter
\def\renewtheorem#1{%
  \expandafter\let\csname#1\endcsname\relax
  \expandafter\let\csname c@#1\endcsname\relax
  \gdef\renewtheorem@envname{#1}
  \renewtheorem@secpar
}
\def\renewtheorem@secpar{\@ifnextchar[{\renewtheorem@numberedlike}{\renewtheorem@nonumberedlike}}
\def\renewtheorem@numberedlike[#1]#2{\newtheorem{\renewtheorem@envname}[#1]{#2}}
\def\renewtheorem@nonumberedlike#1{  
\def\renewtheorem@caption{#1}
\edef\renewtheorem@nowithin{\noexpand\newtheorem{\renewtheorem@envname}{\renewtheorem@caption}}
\renewtheorem@thirdpar
}
\def\renewtheorem@thirdpar{\@ifnextchar[{\renewtheorem@within}{\renewtheorem@nowithin}}
\def\renewtheorem@within[#1]{\renewtheorem@nowithin[#1]}
\makeatother

\renewtheorem{theorem}{Theorem}[section] 
\renewtheorem{lemma}[theorem]{Lemma}
\renewtheorem{corollary}[theorem]{Corollary}
\renewtheorem{definition}[theorem]{Definition}
\newtheorem{fact}[theorem]{Fact}
\newtheorem{assumption}[theorem]{Assumption}
\newtheorem{subobservation}{Observation}[theorem]

\crefname{subobservation}{Observation}{Observations}

\newenvironment{lemmaProof}{\par\noindent\textbf{Proof of Lemma  \lemcnt\space}}{\hfill $\Box_{Lemma ~ \lemcnt}$\smallskip}
\newenvironment{obsProof}{\par\noindent\textbf{Proof of Observation  \obscnt\space}}{\hfill $\Box$\smallskip}
\newenvironment{lemmaProofSketch}{\par\noindent\textbf{Proof Sketch of Lemma  \lemcnt\space}}{\hfill $\Box_{Lemma ~ \lemcnt}$}

\newcommand{\clmcnt}{0}
\newcommand{\lemcnt}{0}
\newcommand{\obscnt}{0}
\newcommand{\thmcnt}{0}

\newcommand{\remove}[1]{}
\newcommand{\reduce}[1]{}
\newcommand{\compact}[1]{}
\newcommand{\eps}{\varepsilon}
\newcommand{\sP}{\mathcal{P}\xspace}

\newcommand{\F}{\mathbb{F}}
\newcommand{\ECC}{\textsf{ECC}}
\newcommand{\ECCthreshold}{\ensuremath{k}\xspace}

\newcommand{\dist}{\Delta}

\newcommand{\Dadv}{d\xspace} 

\newcommand{\true}{\ensuremath{\mathsf{True}}\xspace}

\newcommand{\false}{\ensuremath{\mathsf{False}}\xspace}

\newcommand{\AFRT}{AFRT\xspace}

\newcommand{\MBRB}{MBRB\xspace}
\newcommand{\broadcast}{\textsf{broadcast}\xspace}

\newcommand{\comm}{\textsf{comm}\xspace}
\newcommand{\commP}{\textsf{comm($\cdot$)}\xspace}
\newcommand{\store}{\textsf{store}\xspace}
\newcommand{\psender}{\ensuremath{p_s}\xspace}

\newcommand{\valid}{\textsf{valid}}
\newcommand{\invalid}{\textsf{invalid}}

\crefname{line}{line}{lines}

\patchcmd{\SetProgSty}{ArgSty}{ProgSty}{}{}

\newcommand{\send}{{\textsc{send}}\xspace}
\newcommand{\forward}{{\textsc{forward}}\xspace}
\newcommand{\bundle}{{\textsc{bundle}}\xspace}
\newcommand{\sig}{\mathit{sig}\xspace}
\newcommand{\sigsset}{\mathit{sigs}\xspace}
\newcommand{\fragsset}{\mathit{fragsSet}\xspace}
\newcommand{\frag}{\mathit{fragtuple}\xspace}

\newif\ifannote

\ifannote
    \newcommand{\anninsert}[2]{{\color{#1}#2}}
    \newcommand{\anncomment}[3]{
        {\color{#1}\colorbox{#1}{\bfseries\sffamily\tiny\textcolor{white}{#2}}
        $\blacktriangleright$ \em #3 $\blacktriangleleft$}
    }
\else
    \newcommand{\anninsert}[2]{#2}
    \newcommand{\anncomment}[3]{}
\fi

\newcommand{\rane}[1]{\anninsert{blue}{#1}}

\newcommand{\ft}[1]{\anninsert{black!50!red}{#1}}
\newcommand{\ta}[1]{\anninsert{magenta}{#1}}

\SetKwProg{ForAll}{for all}{ do}{end}
\SetKwProg{ForEach}{for each}{ do}{end}
\SetKwProg{Upon}{Upon }{ do}{end~upon}
\SetKwProg{When}{When }{ do}{end~when}
\SetKwProg{Function}{Function}{ is}{end~function}
\SetKwProg{InitStep}{Init:}{}{end~init}
\SetProgSty{textnormal}
\SetArgSty{textnormal}
\SetKwComment{Comment}{$\rhd$}{}
\SetCommentSty{textit}
\newcommand{\return}{\textsf{return}\xspace}

\DontPrintSemicolon

\newcommand{\computeFragmentMerkleTree}{\textsc{computeFragMerkleTree}}
\newcommand{\isValid}{\textsc{isValid}\xspace}
\newcommand{\getThreshSig}{\textsc{getThreshSig}\xspace}
\newcommand{\isThreshSig}{\ensuremath{\mathit{isThreshSig}}\xspace}
\newcommand{\computeFragCom}{\textsc{computeFragVecCommit}\xspace}
\newcommand{\Commitment}{\ensuremath{\mathsf{Commitment}}\xspace}

\newcommand{\B}{\vspace*{-\smallskipamount}}
\newcommand{\BB}{\vspace*{-\medskipamount}}
\newcommand{\BBB}{\vspace*{-\bigskipamount}}

\newcommand{\Paragraph}[1]{\smallskip\noindent \emph{#1}.}
\newcommand{\Subparagraph}[1]{\smallskip\noindent \emph{{#1}.}}

\newcommand{\Section}[1]{\compact{\B}\section{#1}}
\newcommand{\Subsection}[1]{\smallskip\noindent \textbf{#1.}~~~}

\makeatletter
\@ifpackageloaded{algorithm2e}{%
\PackageInfo{cleveref}{`algorithm2e' support loaded}%
\crefalias{algocf}{algorithm}%
\crefalias{algocfline}{line}%
\crefalias{AlgoLine}{line}%
\let\cref@old@algocf@nl@sethref\algocf@nl@sethref%
\renewcommand{\algocf@nl@sethref}[1]{%
    \cref@old@algocf@nl@sethref{#1}%
    \cref@constructprefix{AlgoLine}{\cref@result}%
    \@ifundefined{cref@AlgoLine@alias}%
        {\def\@tempa{AlgoLine}}%
        {\def\@tempa{\csname cref@AlgoLine@alias\endcsname}}%
    \xdef\cref@currentlabel{%
        [\@tempa][\arabic{AlgoLine}][\cref@result]%
        \csname p@AlgoLine\endcsname\csname theAlgoLine\endcsname}}%
}{}
\makeatother

\newcommand{\algSize}{small} 

\author{Timoth\'e Albouy}{Univ Rennes, Inria, CNRS, IRISA, 35042 Rennes-cedex, France}{timothe.albouy@irisa.fr}{https://orcid.org/0000-0001-9419-6646}{}

\author{Davide Frey}{Univ Rennes, Inria, CNRS, IRISA, 35042 Rennes-cedex, France}{davide.frey@inria.fr}{https://orcid.org/0000-0002-6730-5744}{}

\author{Ran Gelles}{Bar-Ilan University, Israel}{ran.gelles@biu.ac.il}{https://orcid.org/0000-0003-3615-3239}{}

\author{Carmit Hazay}{Bar-Ilan University, Israel}{carmit.hazay@biu.ac.il}{https://orcid.org/0000-0002-8951-5099}{}

\author{Michel Raynal}{Univ Rennes, Inria, CNRS, IRISA, 35042 Rennes-cedex, France}{michel.raynal@irisa.fr}{https://orcid.org/0000-0002-3355-8719}{}

\author{Elad Michael Schiller}{Chalmers University of Technology, Sweden}{elad.schiller@chalmers.se}{https://orcid.org/0000-0003-3258-3696}{}

\author{Fran\c{c}ois Ta\"{i}ani}{Univ Rennes, Inria, CNRS, IRISA, 35042 Rennes-cedex, France}{francois.taiani@irisa.fr}{https://orcid.org/0000-0002-9692-5678}{}

\author{Vassilis Zikas}{Purdue University, USA}{vassilis.zikas@gmail.com}{https://orcid.org/0000-0002-5422-7572}{}

\authorrunning{Albouy, Frey, Gelles, Hazay, Raynal, Schiller, Taïani, and Zikas}

\title{Near-Optimal Communication Byzantine Reliable Broadcast under a Message Adversary}

\titlerunning{Near-Optimal Communication BRB under a Message Adversary}

\keywords{
Asynchronous message-passing,
Byzantine fault-tolerance, 
Message adversary, 
Reliable broadcast with delivery guarantees, 
Erasure-correction codes, 
\ta{Threshold} signatures, 
\ta{Vector commitments}.
}

\ccsdesc[100]{Theory of computation $\rightarrow$ Distributed algorithms}

\Copyright{Timothé Albouy, Davide Frey, Ran Gelles, Carmit Hazay, Michel Raynal, Elad M. Schiller, François Taïani, and Vassilis Zikas}

\acknowledgements{R. Gelles would like to thank 
Paderborn University and CISPA---Helmholtz Center for Information Security 
for hosting him while part of this research was done. %
}

\funding{This work was partially supported by the French ANR projects ByBloS (ANR-20-CE25-0002-01) and PriCLeSS (ANR-10-LABX-07-81), devoted to the modular design of building blocks for large-scale Byzantine-tolerant multi-users applications. Research supported in part by the United States-Israel Binational Science Foundation (BSF) through Grant No.\@ 2020277.}

\nolinenumbers

\begin{document}

\date{} 

\maketitle  

\begin{abstract}
We address the problem of Reliable Broadcast in asynchronous message-passing systems with $n$ nodes, of which up to $t$ are malicious (faulty), in addition to a \emph{message adversary} that can drop some of the messages sent by correct (non-faulty) nodes.
We present a Message-Adversary-Tolerant Byzantine Reliable Broadcast (MBRB) algorithm that communicates ${\cal O}(|m|+n\kappa)$ bits per node, where $|m|$ represents the length of the application message and $\kappa=\Omega(\log n)$ is a security parameter.
This communication complexity is optimal up to the parameter $\kappa$.
This significantly improves upon the state-of-the-art MBRB solution (Albouy, Frey, Raynal, and Taïani, TCS 2023), which incurs communication of ${\cal O}(n|m|+n^2\kappa)$ bits per node.
Our solution sends at most $4n^2$ messages overall, which is asymptotically optimal.
Reduced communication is achieved by employing coding techniques that replace the need for all nodes to (re-)broadcast the entire application message~$m$.
Instead, nodes forward authenticated fragments of the encoding of $m$ using an erasure-correcting code.
Under the cryptographic assumptions of threshold signatures and vector commitments, and assuming $n > 3t+2d$, where the adversary drops at most~$d$ messages per broadcast, our algorithm allows at least $\ell = n - t - (1 + \epsilon)d$ (for any arbitrarily low $\epsilon> 0$) correct nodes to reconstruct $m$, despite missing fragments caused by the malicious nodes and the message adversary.
\end{abstract}

\widowpenalty=10000
\clubpenalty=7500

\Section{Introduction}
\emph{Reliable Broadcast} allows $n$ asynchronous nodes to agree eventually on a message sent by a designated node, the \emph{sender}, despite the possible malicious behavior by some nodes and the transmission network. 
Reliable broadcast plays a crucial role in key applications, including consensus algorithms, replication, event notification, and distributed file systems.
These systems sometimes require broadcasting large messages or files (\eg permissioned blockchains), and thus, reducing the communication overhead to a minimum is an important aspect of achieving scalability.
In that vein, this work aims at providing \emph{communication efficient} solutions for the task of reliable broadcast in the presence of node and link faults.

%

\emph{Byzantine} nodes~\cite{Dolev82,LSP82} are faulty nodes that are assumed to act cooperatively in an arbitrary manner to hinder the non-faulty nodes (also known as correct nodes) from reaching an agreement on the value of a sent message.
These faulty nodes can manifest in various ways, such as delivering fake messages that were never sent, altering the payload of messages sent by any faulty nodes, delaying message delivery, or even omitting messages altogether. Also, a Byzantine failure can present itself differently to different nodes.

Solving reliable broadcast in the presence of Byzantine nodes (a problem known as BRB for Byzantine Reliable Broadcast~\cite{B87}) has been extensively studied for at least four decades. Bracha~\cite{DBLP:conf/podc/BrachaT83,DBLP:journals/jacm/BrachaT85} in particular showed that BRB could be implemented in asynchronous networks as soon as the number $t$ of Byzantine nodes is limited to be less than a third of the nodes. This seminal result has since been followed by hundreds of works, with a various range of optimizations and tradeoffs between different parameters such as resilience, efficiency, and communication; see~\cite{DBLP:books/sp/Raynal18} for an excellent book on this topic.

A significant challenge to reliable broadcast algorithms arises when the message-passing system is unreliable and possibly cooperates with the Byzantine nodes. Link faults~\cite{SW89,DBLP:journals/tcs/SantoroW07} give Byzantine nodes (potentially limited) control over certain network links, enabling them to omit or corrupt messages (an ability captured under the umbrella term \emph{message adversary}~\cite{DBLP:reference/algo/Raynal16a}).
This work focuses on a specific type of \emph{message adversary}~\cite{DBLP:reference/algo/Raynal16a} that can only omit messages sent by correct nodes, but that cannot alter their content. 
This message adversary abstracts cases related to \emph{silent churn}, where nodes may voluntarily or involuntarily disconnect from the network without explicitly notifying other nodes. 
During disconnection, the adversary causes correct nodes to pause the execution of their algorithm temporarily and resume upon reconnecting. 
In the message adversary model, correct nodes may miss messages sent over reliable communication channels by other nodes while they are disconnected, as there is no explicit notification about the message omission.


\Subsection{Problem overview}
\label{sec:probDesc}
%
We assume $n$ nodes over an asynchronous network, where a message can be delayed for an arbitrary yet finite amount of time (unless omitted by the message adversary). 
We assume the existence of $t$ Byzantine nodes and a message adversary capable of omitting up to~$d$ messages per node's broadcast. 
To be more precise, a node communicates through a $\comm$ primitive (or a similar multicast/unicast primitive that targets a dynamically defined subset of nodes),
which results in the transmission of~$n$ messages, with each node being sent one message, including the sender.
The message adversary can omit messages in transit to a subset of at most $d$ correct nodes.
The adversary is only limited by the size of that subset.
For instance, between different $\comm$ invocations, the adversary can modify the set of correct nodes to which messages are omitted.
Furthermore, a designated sender node holds a message $m$ that it wishes to broadcast to all the nodes. 

An algorithm that satisfies the requirements of reliable broadcast despite Byzantine nodes and a message adversary is called a \emph{Message-adversary Byzantine Reliable Broadcast} (\MBRB) algorithm.
The detailed version of \MBRB's requirements was formulated in~\cite{AFRT23}, see \cref{sec:MBRBdef}.
We informally summarize them here as follows.
\textbf{(1)} For any sender invoking the broadcast algorithm, no two correct nodes deliver $m$ and $m'$, such that $m'\ne m$.
\textbf{(2)} 
For any sender invoking the broadcast algorithm,
either zero or at least $\ell$ correct nodes will deliver~$m$.
The quantity $\ell$ might depend on the adversary's power, \ie on $t$ and~$d$.
\textbf{(3)} If a correct node delivers some message~$m$ from a \emph{correct sender}, this correct sender has broadcast $m$ previously and at least $\ell$ correct nodes will deliver it.

\Subsection{Background}
\label{sec:backGround}
Albouy, Frey, Raynal, and Taïani~\cite{AFRT23} recently proposed a Message-adversary Byzantine Reliable Broadcast algorithm (denoted \AFRT for short) for asynchronous networks that withstands the presence of $t$ Byzantine nodes and a message adversary capable of omitting up to $d$ messages per node's broadcast.
\AFRT guarantees the reliable delivery of any message when $n>3$t+$2d$. 
Moreover, they demonstrate the necessity of this bound on the number of Byzantine nodes and omitted messages, as no reliable broadcast algorithm exists otherwise.


One caveat of \AFRT regards its communication efficiency. 
While it achieves an optimal number of $\bigO{n^2}$ messages, and an optimal delivery power $\ell=n-t-d$, each node's communication requires $\bigO{n\cdot(|m|+n\kappa)}$ bits, where $|m|$ represents the number of bits in the broadcast message 
and $\kappa$ is the length of the digital signatures used in their algorithm. 
In the current work, we design an algorithm that significantly reduces the communication cost per node while preserving the total number of messages communicated.  
Our solution features at most~$4n$ messages per correct node (corresponding to $4n^2$ messages overall), and only \ta{$\bigO{|m| + n\kappa}$ bits per correct node.}
Overall, \ta{$\bigO{n|m| + n^2\kappa}$} bits are communicated by correct nodes.
\ta{This bound is tight (up to the size of the signature~$\kappa$) for deterministic algorithms using signatures~\cite{DBLP:conf/ccs/DasX021,nayak_et_al:LIPIcs:2020:13106}, as every correct node must receive the message~$m$, and as the reliable broadcast of a single bit necessitates at least $\Omega$($n^2$) messages~\cite{DR85}.}

%

\Subsection{Contributions and techniques}
This paper is the first to present an \MBRB algorithm able to tolerate a hybrid adversary combining $t$ Byzantine nodes and a Message Adversary of power $d$, while providing optimal Byzantine resilience, near-optimal communication, and near-optimal delivery power~$\ell$.
\B\begin{theorem}[Main, informal]
\label{thm:main-informal}
For any~$\eps>0$, there exists an efficient \MBRB algorithm, such that every message~$m$ broadcast via this scheme is delivered by at least $\ell = n-t-(1+\eps)d$ correct nodes under the assumption $n>3t+2d$. 
Each correct node communicates no more than $4n$ messages and 
\ta{$\bigO{|m|+n\kappa}$} bits overall, where $|m|$ is the length of the message~$m$. 
\end{theorem}
\B

The above asymptotic communication complexity holds assuming a sufficiently long message~$m$. 
Further, $n-t-\Dadv$ is a natural upper bound on the delivery power~$\ell$ of any MBRB algorithm. This bound arises from the power of the message adversary to isolate a subset of the correct parties of size~$\Dadv$, and omit all messages sent to this subset. 
Our solution obtains a delivery power~$\ell$  that is as close to the limit as desired, at the cost of increasing communication (through the hidden constants in the asymptotic $\bigO{\cdot}$~term, which depends on~$\eps$).
Finally, $n>3t+2d$ is a necessary condition to implement MBRB under asynchrony~\cite{DBLP:conf/opodis/AlbouyFRT22}, thus making our solution optimal in terms of Byzantine resilience.


The starting point of our algorithm is the AFRT algorithm~\cite{AFRT23}. This algorithm achieves all the desired MBRB properties (Definition~\ref{def:mbrb}), albeit with a large communication cost of at least $n^2|m|$ bits overall. 
This communication cost stems from the re-emission strategy used by AFRT.
In the AFRT algorithm, the sender first disseminates the message $m$ to all nodes.
To counter a possibly faulty sender, each node that receives $m$ signs it and forwards it to the entire network, along with its own signature and any other signature observed so far for that message. This re-broadcasting step leads to  $n^2|m|$ bits of communication.


In order to reduce the communication costs,  we apply a coding technique, inspired by an approach by Alhaddad \etal~\cite{DBLP:conf/podc/AlhaddadDD0VXZ22}. Instead of communicating the message~$m$ directly, the sender first encodes the message using an error-correction code and ``splits'' the resulting codeword between the nodes, so that each node receives one fragment of size $\bigO{|m|/n}$ bits.
Now, each node needs to broadcast only its fragment of the message rather than the entire message. This reduced per-node communication effectively reduces the overall communication for disseminating the message itself to $n|m|$ bits.

Due to the message adversary and the actions of the Byzantine nodes, some of the fragments might not arrive at their destination.
Error-correction codes have the property that the message~$m$ can be reconstructed from any sufficiently large subset of the fragments.
But Byzantine nodes can do even worse, namely, they can propagate an incorrect fragment.
Correct nodes cannot distinguish correct fragments from incorrect ones (at least, not until enough fragments are collected, and the message is reconstructed).
Without this knowledge, correct nodes might assist the Byzantine nodes in propagating incorrect fragments, possibly harming the correctness and/or performance of the algorithm. 
To prevent this, the sender could sign each fragment that it sends.
A node receiving a fragment could then verify that it is correctly signed by the sender and ignore it otherwise.
The drawback of this solution is that only the sender can generate signatures for fragments.

In our MBRB algorithm, we rely on correct nodes that have already reconstructed the correct message to disseminate its fragments to the nodes that have not received any (say, due to the message adversary).
In principle, when a node reconstructs the correct
message, it can generate the codeword and obtain all the fragments, even if it did not receive some of them beforehand.
However, the node cannot generate the sender's signature for the fragments it generated by itself.
Because of this, the node cannot relay these fragments to the other nodes, potentially leading to a reduced delivery power~$\ell$.  

\rane{We avert this issue by exploiting vector commitments~\cite{CF13}. This cryptographic primitive generates a unique short digest~$C$ for any input vector of elements~$V$.
Additionally, it generates succinct proofs of inclusion for each element in~$V$.
%
In our system, the fragments of the (coded) message~$m$ form the vector~$V$, and the inclusion proofs replace the need to sign each fragment separately.
In more detail, every fragment of the codeword communicated by some node is accompanied by two pieces of information: the commitment $C$ for the vector $V$ containing all fragments of~$m$, and a proof of inclusion showing that the specific fragment indeed belongs to~$V$ (see \cref{sec:vc} for a formal definition of \ft{these properties}). 
The sender signs only the commitment~$C$. 
This means that Byzantine nodes cannot generate an incorrect fragment and a proof that will pass the verification, since they cannot forge the sender's signature on~$C$.
Yet, given the message~$m$, generating a proof of inclusion for any specific fragment can be done by any node.
The vector commitment on the message $m$  creates the same commitment~$C$ and the same proofs of inclusion generated by the sender.
These could then be propagated to any other node along with the sender's signature on~$C$.
}


To complete the description of our MBRB algorithm, we mention that, similar to AFRT, our algorithm tries to form a quorum of signatures on some specific \ta{vector commitment~$C$}.
In parallel, nodes collect fragments they verify as part of the message whose \ta{vector commitment is~$C$}. 
Once a node collects enough signatures (for \ta{some~$C$}) and at the same time obtains enough message fragments that are proven to belong to the \ta{same~$C$}, the node can reconstruct~$m$ and deliver (accept) it. 
At this point, the node also disseminates the quorum of signatures along with (some of) the fragments. 
This allows other correct nodes to reconstruct the message and verify a quorum has been reached. 
In fact, the dissemination of fragments, including fragments that this node did not have before reconstructing the message,  is a crucial step in amplifying the number of nodes that  deliver~$m$ to our stated level of $\ell=n-t-(1+\eps)\Dadv$.
See the full description of the MBRB algorithm in \cref{sec:MBRBscheme}.
%
%

Although our algorithm builds quorums on \ta{commitments}, it departs substantially from the BRB algorithm proposed by Das, Xiang, and Ren~\cite{DBLP:conf/ccs/DasX021}, which avoids signatures and relies on hashes only.
Their solution provides an overall communication complexity in \ta{$\bigO{n|m|+n^2\kappa}$} that is optimal up to the $\kappa$ parameter.
Following the sender's initial dissemination of message $m$, their proposal runs Bracha's algorithm on the hash value of the broadcast message to ensure agreement.
Unfortunately, when used with a message adversary, Bracha's algorithm loses the sub-optimal Byzantine resilience $n>3t+2d$ that AFRT and our solution provide, which is why the solution presented in this paper avoids it.
(See \cref{sec:Bracha:with:MA} for a more detailed discussion of why this is so.)
%
For the sake of presentation, our algorithm's complete analysis and proof details appear in the Appendix.

\Subsection{Related work}
\emph{Byzantine reliable broadcast} (BRB) can be traced back to Pease, Shostak, and Lamport~\cite{DBLP:journals/jacm/PeaseSL80}, which considered the particular case of synchronous message-passing systems.
Since then, solutions for reliable broadcast, together with the related \emph{consensus} problem, have been considered for many distributed systems~\cite{DBLP:books/sp/Raynal18}. 
In asynchronous systems, BRB solutions can be traced back to Bracha and Toueg~\cite{DBLP:conf/podc/BrachaT83,DBLP:journals/jacm/BrachaT85}.
Recent advances~\cite{DBLP:conf/srds/MaurerT14,DBLP:conf/opodis/GuerraouiKKPST20,DBLP:conf/icdcs/BonomiDFRT21,DBLP:conf/sss/DuvignauRS22} in Byzantine fault-tolerant (BFT) solutions to the problem of BRB include the above-mentioned \AFRT~\cite{AFRT23}, which safeguards also against the message adversary.
Our solution features substantially lower communication than \AFRT, without harming the other properties, \eg the number of messages or the delivery power.

Computation in networks with link faults, namely, with Byzantine links was considered by Santoro and Widmayer~\cite{SW89,SW90}, who discussed various types of Byzantine actions, \eg omitting messages, corrupting messages, injecting messages, and combinations thereof. 
The works of~\cite{DBLP:journals/tcs/SantoroW07,BDP97} focus on the case of reliable broadcast with such faults.
In~\cite{Pelc92}, Pelc proved that robust communication is feasible over graphs whose edge-connectivity is more than~$2f$, assuming the number of Byzantine links is bounded by~$f$.
This is also implied by the work of Dolev~\cite{Dolev82}. 
Censor-Hillel \etal~\cite{CCGS22} and Frei \etal~\cite{FGGN24} show that any computation can be performed when all links suffer arbitrary substitution faults (but no crashes), given that the network is 2-edge connected. 
When all links suffer corruption, but their overall amount is restricted, any computation can be reliably performed by Hoza and Schulman~\cite{HS16}, for synchronous networks where the topology is known, or by Censor-Hillel, Gelles, and Haeupler~\cite{CGH19}, for asynchronous networks with unknown topology, see also~\cite{gelles17}.

Settings that allow Byzantine nodes in addition to faulty links were considered~\cite{PT86,gong1998byzantine,SCY98,Dasgupta98,biely03,DBLP:journals/tcs/BielySW11}.
Building on~\cite{AFRT23}, the algorithm was also extended to signature-free asynchronous systems~\cite{DBLP:conf/opodis/AlbouyFRT22}, albeit with lower delivery guarantees, and a weaker resilience with respect to~$d$ and~$t$.
We consider fully connected asynchronous networks, Byzantine nodes, and omission only link failures. 
But, our \MBRB is the first, to the best of our knowledge, to offer a near-optimal communication (up to the length of the signature, $\kappa$) and delivery power~$\ell$.



Coding techniques are implemented to minimize the dissemination costs associated with message transmission across the network, ensuring the ability to reconstruct data in the event of node failures or adversarial compromises.
In the context of Blockchains, significant contributions have been made by Cachin and Tessaro~\cite{DBLP:conf/wdag/CachinT05a} as well as Cachin and Poritz in SINTRA~\cite{DBLP:conf/dsn/CachinP02}, followed by its successors such as HoneyBadgersBFT~\cite{DBLP:conf/ccs/MillerXCSS16}, BEAT~\cite{DBLP:conf/ccs/DuanRZ18}, DispersedLedger~\cite{DBLP:conf/nsdi/YangPAKT22} and Alhaddad \etal~\cite{DBLP:conf/podc/AlhaddadDD0VXZ22}.
These solutions leverage digital signatures and coding techniques to provide a balanced and reliable broadcast.
Our work contributes to the advancement of the state of the art in the field of coded reliable broadcast by offering improved fault-tolerance guarantees that are stronger than the aforementioned solutions.





\Section{Preliminaries}
\label{sec:prelim}
\B
\Paragraph{General notations and conventions}
For a positive integer~$n$, let $[n]$ denote the set $\{1,2,\ldots,n\}$.
A sequence of elements $(x_1,\ldots,x_n)$ is shorthanded as $(x_i)_{i\in[n]}$.
We use the symbol~`$\bull$' to indicate any possible value. 
That is, $(h,\bull)$ means a tuple where the second index includes any arbitrary value which we do not care about. 
All logarithms are base 2.
	
\Subsection{Nodes and Network} 
We focus on asynchronous message-passing systems that have no guarantees of communication delay. 
The system consists of a set, $\sP=\{ p_1, \ldots, p_n\}$, of $n$ fail-prone nodes \ft{that cannot access a clock or use timeouts}.
We identify node~$i$ with~$p_i$. 
		
\Paragraph{Communication means} 
Any ordered pair of nodes $p_i,p_j \in \sP$ has access to a communication channel, $\mathit{channel}_{i,j}$.
Each node can send messages to all nodes (possibly by sending a different message to each node).
That is, any node, $p_i\in \sP$, can invoke the transmission macro, $\comm(m_1,\ldots, m_n)$, that communicates the message $m_j$ to~$p_j$ over $\mathit{channel}_{i,j}$.
The message~$m_j$ can also be empty, in which case nothing will be sent to~$p_j$.
However, in our algorithms, all messages sent in a single $\comm$ activation will have the same length.
%
Furthermore, when a node sends the same message $m$ to all nodes, we write $\broadcast(m)=\comm(m,m,\ldots,m)$ for shorthand.
We call each message $m_j$ transmitted by the protocol an \emph{implementation message} (or simply, a \emph{message}) to distinguish such messages from the \emph{application}-level messages, \ie the one the sender wishes to broadcast.

\Paragraph{Byzantine nodes} 
\label{sec:nodeFailures}
Faulty nodes are called \emph{Byzantine}, and their adversarial behavior can deviate from the proposed algorithm in any manner.
For example, they may crash 
or send fake messages. 
Their ability to communicate and collude is unlimited.
They might perform any arbitrary computation, and we assume their computing power is at least as strong as that of non-faulty nodes, yet not as strong as to undermine the security of the cryptographic signatures we use.
We assume that at most $t$ nodes are faulty, where $t$ is a value known to the nodes.  
Non-faulty nodes are called \emph{correct nodes}. 
The set of correct nodes
contains $c$ nodes where $n-t \leq c \leq n$. 
The value of $c$ is unknown.


 
Faulty nodes may deviate arbitrarily from the correct implementation of $\commP$. For instance, they may unicast messages to only a subset of the nodes in~$\sP$.
As mentioned, each pair of nodes can communicate  using~$\mathit{channel}_{i,j}$.
While $\mathit{channel}_{i,j}$ is assumed to be a reliable channel that is not prone to message corruption, duplication, or the creation of fake messages that were never sent by nodes;
the \emph{message adversary}~\cite{SW89,DBLP:journals/tcs/SantoroW07,DBLP:reference/algo/Raynal16a}, which we specify next, has a limited ability to cause message loss.

\Paragraph{Message adversary} 
\label{sec:messageAdversary}
This entity can remove implementation messages from the communication channels used by correct nodes when they invoke~$\commP$. 
More precisely, during each activation of $\comm(m_1,\ldots,m_n)$, the adversary has the discretion to choose up to ${\Dadv}$ messages from the set~$\{m_i\}$ 
and eliminate them from the corresponding communication channels where they were queued. We assume that the adversary has full knowledge of the contents of all messages~$\{m_i\}$, and thus it makes a worst-case decision as to which messages to eliminate.

The failures injected by a message adversary differ from those of classical sender and/or receiver omissions in that they are \emph{mobile}.
\Ie they are not pinned to a set of particular nodes but may move between any correct nodes during the same execution since they are defined per invocation of~$\commP$. 
In particular, no node is immune to the message adversary.

Note that for the case of ${\Dadv} = 0$, the adversary is as weak as the common settings in which all communication channels are reliable (since no message is ever lost) and $t$ nodes are Byzantine.
Assumption~\ref{asm:MBRBassumption} limits the adversary's power to avoid network partitions.
As mentioned above, this is necessary for any MBRB algorithm~\cite{AFRT23}. 

\B
\begin{assumption}[Adversary-power-assumption]
\label{asm:MBRBassumption}
$n > 3t$+$2{\Dadv}$.
\end{assumption}	
\B
Since the message adversary can omit all implementation messages that are sent to a given set $D\subseteq \sP:|D|={\Dadv}$, we know that $\ell$, the number of correct nodes that are guaranteed to output the broadcast~$m$ correctly, must satisfy $\ell \leq c-{\Dadv}$. 

Such a hybrid fault model has been \ft{studied} in the past in synchronous networks~\cite{DBLP:journals/tcs/BielySW11}, but has remained little studied in an asynchronous setting, except for the work of Schmid and Fetzer~\cite{DBLP:conf/srds/SchmidF03} that limits itself to round-based algorithms (unlike us) and does not cover full disconnections of correct nodes (which we do).
Modeling full disconnections is relevant 
as this captures correct nodes that remain disconnected for long periods.
In addition to bounding the maximal number of outgoing messages that a correct sender might lose (as we do), the model proposed by Schmid and Fetzer~\cite{DBLP:conf/srds/SchmidF03} also bounds the maximal number of incoming messages that any correct node might miss.
This adds an elegant symmetry to the model but poses significant challenges when considering asynchronous networks, as there is no obvious scope
on which to limit the number of incoming messages missed by a given node. Schmid and Fetzer~\cite{DBLP:conf/srds/SchmidF03} \ft{therefore restrict} the fault model to round-based algorithms. \ft{By contrast, our model allows} 
\ft{for} algorithms that do not follow this structure.


\newcommand{\nECC}{{\bar n}}
\newcommand{\kECC}{{\bar k}}
\newcommand{\rECC}{{\bar r}}
\newcommand{\dECC}{{\bar d}}

\Subsection{Error Correction Codes} 
\label{sec:ecc}
A central tool used in our algorithm is an error-correction code (ECC)~\cite{roth06}. Intuitively speaking, an ECC takes a message as input and adds redundancy to create a codeword from which the original message can be recovered
even when parts of the codeword are corrupted. 
In this work, we focus on \emph{erasures}, a corruption that replaces a symbol of the codeword with a special erasure mark~$\bot$.

Let $\F$ denote a finite field whose size we set later, and let $\bot$ be a special symbol~$\bot \notin \F$. 
Given two strings of the same length, $x,y \in \F^n$, their \emph{Hamming distance} is the number of indices where they differ, $\Delta(x,y) = | \{i\in[n] \mid x_i \ne y_i\}|$. Given a subset $I \subseteq [n]$, we denote by $x_I \in \F^{|I|}$ the string $x$ restricted to the indices in~$I$.

To avoid confusion with global parameters, we denote the ECC-specific parameters by using a bar (\eg $\bar x$).
An \emph{error-correction code} is a function $\ECC : \F^\kECC \to \F^\nECC$, with \emph{rate}  $\rECC=\kECC/\nECC$, and \emph{distance} ${\dECC} =  \min_{x,y\in \F^\kECC, x\ne y} \dist( \ECC(x), \ECC(y))$.
%
The Singleton bound determines that $\dECC\le \nECC-\kECC+1$, and when the equality holds, 
the code is said to be maximum distance separable (MDS).
A prominent example of MDS codes is Reed-Solomon (RS) codes~\cite{RS60}, which exist for any $\kECC,\nECC$, and $|\F|\ge \nECC$. Such codes can be efficiently encoded and decoded~\cite{roth06}.
The erasure correction capabilities of a code depend on its distance, as given by the following fact.


\B
\begin{fact}[Erasure Correction Capability]
    \label{fct:ErasureECC}
    Any error-correction code of distance ${\dECC}$ can recover up to ${\dECC}-1$ erasures. 
    That is, for any $y\in (\F\cup \{\bot\})^\nECC$, let $E = \{ i \mid y_i = \bot\}$ the set of erased indices. Then, if $|E|<{\dECC}$, there is at most a single $x\in \F^\kECC$ such that 
    $y_{[\nECC]\setminus E}=\ECC(x)_{[\nECC]\setminus E}$.  
\end{fact}
\B

\Subsection{Cryptographic Primitives} \label{sec:crypto-prim}
Our algorithm relies on cryptographic assumptions.
We assume that the Byzantine nodes are computationally bounded with respect to the security parameter, denoted by~$\kappa$.
That is, all cryptographic algorithms are polynomially bounded in the input~$1^\kappa$.
We assume that $\kappa =\Omega$($\log n$).
We further assume the availability of Public Key Infrastructure (PKI), which is the setting that assumes each node is assigned a pair of public/private keys generated according to some standard key-generation algorithm.
Further, at the start of the computation, each node holds its own private key and the public key of all other parties.
This setting implies private and authenticated channels.
In particular, each node has public and private keys to support the following cryptographic primitives.


\ta{\Subparagraph{Threshold signatures}}
\label{sec:cryptographicSignatures}\label{sec:signatures} 	
\newcommand{\TSIG}{\ensuremath{\mathsf{TSIG}}\xspace}
\newcommand{\TSIGN}{\ensuremath{\mathsf{ts\_sign\_share}}\xspace}
\newcommand{\TVERS}{\ensuremath{\mathsf{ts\_verify\_share}}\xspace}
\newcommand{\TCOMB}{\ensuremath{\mathsf{ts\_combine}}\xspace}
\newcommand{\TVER}{\ensuremath{\mathsf{ts\_verify}}\xspace}
\ta{In a \emph{$(\tau,n)$ threshold signature} scheme~\cite{S00}, at least $\tau$ out of all $n$ nodes (the threshold) produce individual \emph{signatures shares} $\sigma$ for the same message $m$, which are then aggregated into a fixed-size \emph{threshold signature} $\Sigma$.
\ft{Verifying $\Sigma$ does not require the public keys of the signers}; one needs to use a single system-wide public key, the same for all threshold signatures produced by the scheme.
This system public key, known to everyone, is generated during the system setup phase and distributed through the PKI.

Formally, we define a $(\tau,n)$ threshold signature scheme as a tuple of (possibly randomized) algorithms $\TSIG = (\TSIGN,\TVERS,\TCOMB,\TVER)$.
The signing algorithm executed by node~$p_i$ (denoted, $\TSIGN_i$) takes a message~$m$ (and implicitly a private key) and produces a signature $\sigma = \TSIGN_i(m)$.
The share verification algorithm takes a message $m$, a signature share $\sigma_i$, and the identity $i$ of its signer $p_i$ (and implicitly $p_i$'s public key), and outputs a single bit, $b=\TVERS(m,\sigma_i,i) \in \{\valid,\invalid\}$, which indicates whether the signature share is valid or not.
The combination algorithm takes a set $\sigsset$ of $\tau$ valid signature shares produced by $\tau$ out of $n$ nodes and the associated message $m$ (and implicitly the system public key) and outputs a threshold signature $\Sigma = \TCOMB(\sigsset)$.
The threshold signature verification algorithm takes a message $m$ and a threshold signature $\Sigma$ (and implicitly the system public key) and outputs a single bit $b=\TVER(m,\Sigma) \in \{\valid,\invalid\}$, indicating if the threshold signature is valid or not.

We require the conventional robustness and unforgeability properties for threshold signatures. 
This scheme is parameterized by a security parameter~$\kappa$, and the size of signature shares and threshold signatures, $|\sigma|=|\Sigma|=\bigO{\kappa}$, is independent of the size of the signed message, $m$.
In our algorithm, we take $\tau=\lfloor\frac{n+t}{2}\rfloor{+}1$ (\ie the integer right above $\frac{n+t}{2}$).
}

\newcommand{\merkleFragments}{n} 

\ta{\Subparagraph{Vector commitments (VC)}}
\label{sec:vc}
\newcommand{\vcCommit}{\ensuremath{\mathsf{vc\_commit}}\xspace}
\newcommand{\vcVerify}{\ensuremath{\mathsf{vc\_verify}}\xspace}
\ta{A \emph{vector commitment} (VC) is a short digest $C$ for a vector of elements $V$, upon which a user can then generate a \emph{proof of inclusion} $\pi$ (sometimes called \emph{partial opening}) of some element in $V$ without disclosing the other elements of $V$ to the verifier: the verifier only needs $C$, $\pi$, the element, and its index in the vector to verify its inclusion in $V$.
A Merkle tree~\cite{Merkle90} is a notable example of vector commitment, although with several sub-optimal properties.
For example, for a hash size of $\kappa$, a Merkle proof of inclusion is of $\bigO{\kappa\log |V|}$ bits, which is significantly larger than modern schemes such as Catalano-Fiore vector commitments~\cite{CF13}, which produce proofs of inclusion with an optimal size of $\bigO{\kappa}$ bits.
In our construction, we use these optimal VC schemes (such as Catalano-Fiore's), which provide commitments and proofs in $\bigO{\kappa}$ bits.
The VC scheme provides two operations, parameterized by the security parameter~$\kappa$: $\vcCommit(\cdot)$ and $\vcVerify(\cdot)$, that work as follows.
For any vector of strings $V=(x_1,\ldots,x_n) = (x_i)_{i\in[n]}$, the function $\vcCommit(V) \rightarrow (C,\pi_1,\ldots,\pi_n)$ returns $C$, the commitment, and every $\pi_i$, the proof of inclusion for $x_i$.
The following hold.
\begin{enumerate}
    \item \textbf{Proof of inclusion (Correctness):}
    Let $(C,(\pi_i)_{i\in[n]}) = \vcCommit((x_1,\ldots,x_n))$.
    Then for any $i\in[n]$, it holds that $\vcVerify(C,\pi_i,x_i,i)=\valid$.
    \item \textbf{Collision-resistance (Binding):}
    \sloppy
    For any $j\in[n]$ and any randomized algorithm $A$ \ft{taking} $(x_i)_{i\in[n]}$ and $(C, (\pi_i)_{i\in[n]}) = \vcCommit(x_1,\ldots, x_n)$ as input, 
    $\Pr[A \text{ outputs } (x_j',\pi'_j,j) \wedge  \vcVerify(C,\pi'_j,x_j',j)=\valid] < 2^{-\Omega\text{(}\kappa\text{)}}$.
    Namely, it is difficult to generate $x_j'\ne x_j$ \ft{and} a \ft{valid} proof~$\pi_j'$ 
    \ft{for} the same \ft{commitment}~$C$.
\end{enumerate}
We omit the traditional Hiding property of VC schemes (a commitment does not leak information on the vector~\cite{AAFGRRT24}) as it is unneeded in our algorithm.
We also implicitly assume that the \vcCommit operation is deterministic: it always returns the same commitment $C$ given the same input vector $V$, no matter the calling process $p_i$.
This is the case for Catalano-Fiore's scheme~\cite{CF13}, which does not use random salt. 
}

\Subsection{Specification of the MBRB primitive}
\label{sec:MBRBdef}
\MBRB's objective is to guarantee a reliable broadcast, meaning it aims to ensure that a bounded minimum number of correct nodes ultimately deliver the broadcast messages to the application while upholding specific safety and liveness criteria. This assurance holds even when confronted with Byzantine faults and a message adversary capable of selectively suppressing messages.


An MBRB algorithm \ft{provides} the MBRB-broadcast and MBRB-deliver operations.
The following specification is presented in a single-shot, single-sender version, \ft{where  \psender is} the sending node. 
A multi-sender, multi-shot version can be derived by adding the sender's identifier and a running sequence number to messages and signatures.
Nodes invoke the MBRB-deliver operation to deliver (to the application layer) messages broadcast by~\psender.

\begin{table}[t!]
	\centering
	\begin{\algSize}
	\begin{tabular}{|c|l|}
		\hline
		$n$ & total number of nodes\\
		$t$ & upper bound on the number of Byzantine nodes\\
		$d$ & removal power of the message adversary\\
		$\ECCthreshold$ & reconstruction threshold of the erasure code, $\ECCthreshold$ out of $n$\\
		$\psender$ & the designated sending node (with identity $s$)\\
		$\sigma_i$ & signature \ta{share} by node $p_i$\\
		$\sig_i$    & the pair $(\sigma_i,i)$\\
		$\sigsset, \sigsset_i$ & sets of (signature \ta{share},id) pairs\\
            \ta{$\Sigma$} & \ta{threshold signature (TS)} \\
            \ta{$\tau$} & \ta{threshold of the TS scheme (set to $\tau{=}\lfloor\frac{n+t}{2}\rfloor{+}1$ in our algorithm)} \\
		$m, m', m_i$ & application messages\\
		\ta{$C, C', C_m$} & \ta{vector commitments} \\
		$\tilde m_i$ & $i^\mathrm{th}$ message fragment of application message $m$\\
		$\tilde \pi_i$ & proof of inclusion of fragment $\tilde m_i$\\
		\hline
	\end{tabular}
	\caption{Notations used by \cref{alg:3.1:merkle:trees}}\label{tab:notations}
	\end{\algSize}
 \compact{\BBB}
\end{table}

Definition~\ref{def:mbrb} specifies the safety and liveness properties of MBRB. 
Safety ensures that messages are delivered correctly without spurious messages, duplication, or duplicity. 
Liveness guarantees that if a correct node broadcasts a message, it will eventually be delivered by at least one correct node (MBRB-Local-delivery). 
If a correct node delivers a message from any specific sender, that message will eventually be delivered by a sufficient number,~$\ell$, of correct nodes (MBRB-Global-delivery), where $\ell$ is a measure of the \emph{delivery power} of the MBRB object.
The parameter $\ell$ might depend on the adversary's power, \ie on $t$ and~$\Dadv$.
Since the message adversary can omit all implementation messages that are sent to an unknown set $D\subseteq \sP:|D|={\Dadv}$, we know that $\ell \leq c-{\Dadv}$.

\compact{\B}
\begin{definition}
	\label{def:mbrb}
	An \MBRB is an algorithm that satisfies the following properties.
	\begin{itemize}
		\item  \textbf{MBRB-Validity.~~} \ft{If} \psender is correct and a correct node, $p_i$, MBRB-delivers an application message~$m$, then, node~\psender has MBRB-broadcast~$m$ (before that MBRB-delivery).
		\item  \textbf{MBRB-No-duplication.~~} A correct node~$p_i$ MBRB-delivers at most one application message~$m$.
		\item  \textbf{MBRB-No-duplicity.~~} No two different correct nodes MBRB-deliver different application messages from node~\psender.
		
		\item \textbf{MBRB-Local-delivery.~~} Suppose \psender is correct and MBRB-broadcasts an application message~$m$. At least one correct node,~$p_j$, eventually MBRB-delivers~$m$ from node~\psender.
		\item \textbf{MBRB-Global-delivery.~~} Suppose a correct node,~$p_i$, MBRB-delivers an application message~$m$ from~\psender.
		Then, at least $\ell$ correct nodes MBRB-deliver~$m$ from~\psender.
	\end{itemize}
\end{definition}


		
		

\Section{The Coded-MBRB algorithm}
\label{sec:MBRBscheme} \label{sec:ECC-instantiation}
The proposed solution, named Coded MBRB (\cref{alg:3.1:merkle:trees}), allows a distinguished sender~$\psender$ \ta{(known to everyone)} to disseminate one specific application message~$m$. 
%
In \cref{sec:multi}, we discuss how to extend this algorithm so that it implements a general MBRB algorithm, allowing any node to be the sender, as well as allowing multiple instances of the MBRB, either with the same or different senders, to run concurrently.
\ta{In the algorithm, we instantiate the threshold signature scheme with the threshold value set to $\tau=\lfloor\frac{n+t}{2}\rfloor+1$ (see \cref{sec:prelim}).}

\cref{alg:3.1:merkle:trees} introduces the \textsc{MBRB-broadcast}($m$) operation, which takes message~$m$ and disseminates it reliably to a minimum bound of correct nodes, denoted $\ell$.
That is, after executing \cref{alg:3.1:merkle:trees}, and assuming a correct sender, at least $\ell$ correct nodes will have invoked the \textsc{MBRB-delivery}$(m)$ procedure, while no correct node will have invoked \textsc{MBRB-delivery} with $m'\ne m$.
\cref{tab:notations} summarizes the notations used by \cref{alg:3.1:merkle:trees}.

\begin{algorithm}[b]
		\begin{\algSize}
    \Function(\ta{\Comment*[f]{Computes ECC fragments and VC}}){\ta{$\computeFragCom(m)$}}{
		$\tilde m \gets \ECC(m)$\label{ln:tildeMgetsECCm}
		\Comment*{See ``The error-correction code in use'' paragraph}
		\textbf{let} $\tilde m_1,\ldots, \tilde m_n$	be $n$ equal size fragments of $\tilde m$\label{ln:letM1ldots:new}\;
		\ta{$(C,\pi_1,\ldots,\pi_n) \gets \vcCommit( \tilde m_1,\ldots, \tilde m_n )$}\label{line:v3.1:MTree}\label{line:v3.1:MHash} ;
		
		\return $\big(\ta{C}, (\tilde m_j, \pi_j ,j )_{j\in[n]} \big)$
    }
    \medskip
    
    \Function(\ta{\Comment*[f]{Checks the validity of received msgs}}){$\isValid(\ta{C},\fragsset,\sigsset\ta{,\isThreshSig})$}{
        \ta{
        \Comment{If $\sigsset$ is a set of signature shares}
        \uIf{$\neg\isThreshSig$}{
            \Comment{Each signature in $\sigsset$ must be valid}
            \lIf{$\exists \, (\sigma_x,x) \in \sigsset: \neg\TVERS(C,\sigma_x,x)$}{\return \false} \label{line:isValid:sigs:are:valid}
		\Comment{$\sigsset$ must contain the sender's signature}
		\lIf{$(\bull,s)\not\in\sigsset$\label{line:isValid:sigs:contain:ps:sign}}{\return \false 
		} 
        }
        \smallskip
        \Comment{If $\sigsset$ is a threshold signature, check if it aggregates at least $\tau{=}\lfloor\frac{n+t}{2}\rfloor{+}1$ valid shares}
        \lElseIf{$\neg\TVER(C,\sigsset)$}{\return \false}
        }
		
        \smallskip
        \Comment{Each proof of inclusion in $\fragsset$ must be valid}
        $\fragsset\gets\fragsset\setminus\{\bot\}$\Comment*{Ignoring $\bot$ values}
        \ta{\lIf{$\exists\, (\tilde m_x, \pi_x,x)\in\fragsset: \neg\vcVerify(C,\pi_x,\tilde m_x,x)$}{\return \false\xspace\hspace{-0.4em} \label{line:isValid:inclusion:proof}}}%
        \return \true
    }
    \medskip

    \ta{
    \Function(\Comment*[f]{Get the TS for $C$ if it exists, $\bot$ otherwise}){$\getThreshSig(C)$}{
        $\sigsset_C \gets \{\text{signatures shares stored by $p_i$ for commitment $C$}\}$\;
        \return $\Sigma_C \gets \left\{%
      \begin{array}{@{}ll}%
        \text{the threshold signature saved for }C & \text{ if it exists} \\
        \text{else, } \TCOMB(\sigsset_C) & \text{ if } |\sigsset_C|>\frac{n+t}{2} \\
        \bot & \text{ otherwise}\label{line:return:TS:from:storage}
      \end{array}\right.$
    }
    }
  
	\caption{\ta{Helper functions of the Coded MBRB Algorithm (code for $p_i$)}}
	\label{algo:helpers}
		\end{\algSize}

\end{algorithm}

\Subsection{Algorithm description}
\label{sec:SchemeDescription}
\textsc{MBRB-broadcast}$(m)$ allows the sender $\psender$ to start disseminating the application message, $m$ (\cref{ln:MBRBbroadcastMsn}).
The sender (\cref{line:fragments:for:m}) \ft{starts by} invoking \ta{$\computeFragCom(m)$} (\cref{algo:helpers}).
This function encodes the message~$m$ using an error-correction code, divides it into $n$ fragments and constructs a \ta{vector commitment with an inclusion proof for \ft{each} fragment.}
The function returns several essential values: \ta{the commitment $C$}, and the fragment details~$(\tilde m_j, \pi_j ,j)$, which contain the fragment data itself~$\tilde m_j$ (the $j$-th part of the codeword $\ECC(m)$; see below for \ft{detail}),
a proof of inclusion~$\pi_j$ for that part, and each fragment's respective index $j$. 
For ease of reference, \ta{let $\Commitment(m)$ represent the commitment $C$ obtained from $\computeFragCom(m)$.} 
This \ta{commitment} serves as a compact representation of the entire message $m$.
The sender node \psender is responsible for signing the computed \ta{commitment~$C$} and generating a \ta{signature share} $\sig_s$ (\cref{line:ps:signs:hash}) which includes \psender's identifier.
The sender then initiates $m$'s propagation by employing the operation \ta{\comm} (\cref{ln:v3.1:commV1ldotsVn}), which sends to each node,~$p_j$, an individual message, $v_j$. 
The message~$v_j$ includes several components: the message type (\textsc{send}), the \ta{commitment $C$}, the $j$-th fragment details $(\tilde m_j, \pi_j, j)$, and the \ta{signature share} $\sig_s$ (\cref{line:ps:signs:hash}) for~\ta{$C$}.
%
%
%



The rest of the algorithm progresses in two phases, which we describe in turn. 
The first phase is responsible for message dissemination, which forwards message fragments received by the sender. The other role of this phase is reaching a quorum of nodes that vouch for the same message.
A node vouches for a single message by signing its \ta{commitment}.
Nodes collect and store \ta{signature shares} until it is evident that sufficiently many nodes agree on the same message.  
The subsequent phase focuses on disseminating the quorum of \ta{signature shares} so that it is observed by at least~$\ell$ correct nodes, and on successfully terminating while ensuring the delivery of the reconstructed message.


\begin{algorithm}[t]

\thisfloatpagestyle{empty}
  \newcommand{\announcePhase}[1]{%
    \medskip%
  
    {\SetKwComment{Comment}{}{}\Comment{\fbox{#1}}}
    \smallskip%
  }

    \Function(\Comment*[f]{only executed by the sender,~\psender}){\textsc{MBRBbroadcast}$(m)$}{
        \label{ln:MBRBbroadcastMsn}
        $\big( \ta{C}, (\tilde m_j, \pi_j ,j)_j \big) \gets \ta{\computeFragCom(m)}$\label{line:fragments:for:m}\;
    
        $\sig_s \gets \big(\ta{\TSIGN_s(C)},s\big)$\label{line:ps:signs:hash}\;
    	
    	  $\comm(v_1,\ldots,v_n)$ \textbf{where} $v_j=\langle \send, \ta{C}, (\tilde m_j, \pi_j), \sig_s \rangle$\label{ln:v3.1:commV1ldotsVn}\label{ln:v3.1:sendFrag}\;
    
      }

  \announcePhase{Phase I: Message dissemination} 
  
  \Upon{\label{line:PhaseI:start}%
    $\langle \send, \ta{C'}, (\tilde m_i, \pi_i,i), \sig_s \rangle$ \textbf{arrival from} \psender}{ \label{ln:sendRx:SEND}
    \SetAlgoVlined
    \lIf(){
      $\neg\isValid\big(\ta{C'},\{(\tilde m_i, \pi_i,i)\},\{\sig_s\}\ta{,\isThreshSig{=}\false}\big)$\label{line:isValid:in:SEND}%
    }{\return}
    \lIf{$p_i$ already \ta{executed \cref{line:v3.1:bcast:forward} or signed some commitment $C'' \neq C'$}\label{line:no:sign:twice:SEND}\label{line:condition:sign:hash:SEND}}{\hspace{-0.02em}\return\hspace{-1em}}

    $\sig_i \gets \big(\ta{\TSIGN_i(C')},i\big)$ ;
    \store $\tilde m_i$, $\sig_s$, and $\sig_i$ for \ta{$C'$}\; \label{line:store:in:SEND} \label{line:sign:a:hash:in:SEND}
    
    \broadcast $\langle \forward, \ta{C'}, (\tilde m_i, \pi_i,i), \{ \sig_s,\sig_i \} \rangle$
    \label{line:v3.1:bcast:forward}\label{line:v3.1:forward:after:send}
  }
  \medskip

  \Upon{%
    $\langle \forward, \ta{C'}, \frag_j, \sigsset_j=\{\sig_s,\sig_j\} \rangle$ \textbf{arrival from} $p_j$\label{line:arrival:FORWARD}
  }{
    \SetAlgoVlined
    \lIf(){
      $\neg\isValid\big(\ta{C'},\{\frag_j\},\sigsset_j\ta{,\isThreshSig{=}\false}\big)$\label{line:isValid:in:FORWARD}%
    }{\return}
    \lIf{$p_i$ already signed \ta{some commitment $C''\neq C'$}\label{line:no:sign:twice:FORWARD}\label{line:condition:sign:hash:FORWARD}}{\return}
    %
    %
    \store $\sigsset_j$ for \ta{$C'$}\label{line:store:in:FORWARD}\;
    \If{$\frag_j\neq\bot$}{
      $(\tilde m_j, \pi_j,j) \gets \frag_j$\label{line:store:my:fragment:in:FORWARD} ;
      \store $\tilde m_j$ for \ta{$C'$}
    }
    \If{no \forward message sent yet\label{line:v3.1:test:forward:sent}}{
      $\sig_i \gets \big(\ta{\TSIGN_i(C')},i\big)$ ;
      \store $\sig_i$ for \ta{$C'$}\label{line:sign:a:hash:in:FORWARD}\;
      \broadcast$\langle \forward, \ta{C'}, \bot, \{ \sig_s,\sig_i \} \rangle$\label{line:v3.1:forward:after:forward}\label{line:PhaseI:end}
    }
  }
  \announcePhase{Phase II: Reaching Quorum and Termination }
  
  \When{$\left\{\begin{array}{l}\exists \ta{C'}: \ta{\getThreshSig(C') \neq \bot} \wedge \big|\{ \text{stored } \tilde m_j \text{ for } \ta{C'} \}\big|\geq k\\\wedge \text{ no message has been MBRB-delivered yet}\end{array}\right\}$%
    \label{line:quorum:sigs:enough:fragments}\label{ln:bulljsnWasNotMBRBdelivered}%
    \label{line:PhaseII:start}}{
    \SetAlgoVlined
    $m_i \gets \ECC^{-1}(\tilde m_1, \ldots, \tilde m_n)$,
    $\left\{%
      \begin{array}{@{}l}%
        \text{where~$\tilde m_j$ are taken from line~\ref{line:quorum:sigs:enough:fragments};}\\
        \text{when a fragment is missing use $\bot$.}%
      \end{array}\right.$%
    \label{line:v3.1:reconstruct:m}\;
    $\big(\ta{C}, (\tilde m'_j, \pi'_j,j )_j \big) \gets \ta{\computeFragCom(m_i)}$\label{line:v3.1:recomputing:frags:with:proofs}

     \ta{\lIf{$C' \neq C$}{\return}} \label{line:hashes:equal:b4:delivery}
     %
    
      $\ta{\Sigma_C \gets \getThreshSig(C)}$\; \label{line:get-tsig}
      $\comm(v_1,\ldots,v_n)$ \textbf{where} $v_j=\langle \bundle, \ta{C}, (\tilde m'_i,\pi'_i,i), (\tilde m'_j,\pi'_j,j), \ta{\Sigma_C} \rangle$\label{line:v3.1:bundle:comm}

      $\textsc{MBRBdeliver}(m_i)$\label{ln:MBRBdeliverMjSn}\;
     
  }
  \medskip
%
%
  \Upon{%
    $\langle \bundle, \ta{C'}, (\tilde m'_j,\pi'_j,j), \frag'_i, \ta{\Sigma} \rangle$ \textbf{arrival from} $p_j$%
    \label{line:arrival:BUNDLE}}{
    \SetAlgoVlined%
    \lIf(){
      $\neg\isValid\big(\ta{C'},\big\{(\tilde m'_j,\pi'_j,j), \frag'_i\big\},\ta{\Sigma,\isThreshSig{=}\true}\big)$\label{line:isValid:in:BUNDLE}%
    }{\return}
    \store $(\tilde m'_j,\pi'_j,j)$ and \ta{$\Sigma$} for \ta{$C'$}\label{line:store:in:BUNDLE}\;
    \If{no \bundle message has been sent yet\label{line:test:if:already:bundle} 
        $\wedge$  $\frag'_i\neq\bot$\label{line:check:frag:not:bot}}{
      $(\tilde m'_i,\pi'_i,i)\gets \frag'_i$\;
      \store $(\tilde m'_i,\pi'_i,i)$ for \ta{$C'$}\label{line:store:my:fragment:in:BUNDLE}\;
      \broadcast $\langle \bundle, \ta{C'}, (\tilde m'_i,\pi'_i,i), \bot, \ta{\Sigma} \rangle $\label{line:v3.1:bundle:with:bot}\label{line:PhaseII:end}
    }
  }
  \caption{Phases of the Coded MBRB Algorithm (code for $p_i$, single-shot, single-sender\ta{, threshold for the TS scheme $\tau{=}\lfloor\frac{n+t}{2}\rfloor{+}1$})}
  \label{alg:3.1:merkle:trees}

\end{algorithm}

\afterpage{\clearpage}

\Subparagraph{Validating message integrity}
%
The validity of the signatures and inclusion proofs are checked each time a new message is received (at \cref{ln:sendRx:SEND,line:arrival:FORWARD,line:arrival:BUNDLE}) using the function \isValid
(\cref{algo:helpers}). 
All message types (\send, \forward, and \bundle) \ft{carry} a \ta{vector commitment ($C$ or $C'$) and up to two message fragments with their inclusion proofs}.
\ta{Moreover, the \send and \forward types contain up to two signature shares for the provided commitment, and the \bundle type contains a threshold signature for the provided commitment.} 
The validation hinges on 
three key criteria.
Every enclosed \ta{signature share or threshold signature} must be valid and correspond to the \ft{accompanying} \ta{commitment}.
\ta{For \send or \forward messages, the signature share from the designated sending node \psender must be \ft{present}.}
All message fragments must contain valid inclusion proofs for the \ft{provided} \ta{commitment}.
Note that $\pi_i$, the proof of inclusion of~$\tilde m_i$, does not need to be signed by \psender, as the \ta{commitment} already is.

\Subparagraph{Phase I: Message dissemination}
This phase facilitates the widespread distribution of the message fragments ${\tilde m_j}$.
Recall that the sender has sent each node a different (encoded) fragment of the message~$m$, however, no node currently holds enough information to retrieve the message~$m$.
The phase also sets the ground for forming a quorum on the message~$m$. 

When a node receives a $\send$ message from the sender, it begins by validating the fragment's authenticity (\cref{line:isValid:in:SEND}). Subsequently, the node forwards this fragment to all other nodes by broadcasting a $\forward$ message (\cref{line:v3.1:forward:after:send}).

Upon receiving a $\langle \send, \ta{C'},$ $(\tilde m_i, \pi_i,i), \sig_s \rangle$ message from \psender, the recipient $p_i$ validates the incoming message (\cref{line:isValid:in:SEND}).
%
$p_i$ \ft{then} determines whether it had previously broadcast a \forward message at \cref{line:v3.1:forward:after:send} or signed a 
\ta{commitment $C''$} from \psender distinct from the currently received \ta{$C'$}, in which case the incoming message is discarded (\cref{line:condition:sign:hash:SEND}). 
Otherwise, $p_i$ proceeds to store the received information (\cref{line:store:in:SEND}), encompassing the fragment $\tilde m_i$ and the associated \ta{signature share} $\sig_s$, linked to the specific \ta{commitment $C'$}. 
We clarify that $p_i$ never stores multiple copies of the same information, \ie all store operations are to be read as adding an item to a set.
Subsequently, $p_i$ generates its own \ta{signature share} $\sig_i$ for the \ta{commitment $C'$}, storing it for later utilization (\cref{line:sign:a:hash:in:SEND}). 
Node $p_i$ then disseminates all the relevant information, by broadcasting the message
$\langle \forward, \ta{C'}, (\tilde m_i, \pi_i,i), \{ \sig_s,\sig_i \} \rangle$.

The broadcast of a $\forward$ message is instrumental in disseminating information for several reasons. First, up to $\Dadv$ nodes might not receive the sender's $\send$ message.
Second, this is the node's way to disseminate its own fragment and \ta{signature share} for that specific~\ta{$C'$}.  

Upon the arrival of a $\langle \forward, \ta{C'}, \frag_j, \sigsset_j \rangle$ message from $p_j$ (\cref{line:arrival:FORWARD}), the recipient $p_i$ validates the incoming message using the \isValid function (\cref{algo:helpers}), discarding invalid messages (\cref{line:isValid:in:FORWARD}). 
As for \send messages, $p_i$ checks if it already signed a message from \psender with a different \ta{commitment $C''$}, in which case it discards the message (\cref{line:no:sign:twice:FORWARD}).
Subsequently, $p_i$ stores the set of \ta{signature shares} $\sigsset_j$ linked to the specific \ta{commitment $C'$} (\cref{line:store:in:FORWARD}) and $\frag_j$ contained in this message, if any (\cref{line:store:my:fragment:in:FORWARD}).
%
Also, $p_i$ assesses whether a \forward message has been previously dispatched.
If it has already done so, there is no reason to re-send it, and the processing ends here.
Otherwise, similar to above, $p_i$ generates its own \ta{signature share} $\sig_i$ for the \ta{commitment $C'$}, and broadcasts the message $\langle \ta{\forward, C'}, \bot, { \sig_s,\sig_i } \rangle$.
Note that, in this case, $p_i$ is unaware of his own fragment (\ie it has not received a $\send$ message, or otherwise it would have already sent a \forward message in \cref{line:v3.1:forward:after:send}); therefore it sends the sentinel value~$\bot$ instead.

\Subparagraph{Phase II: Reaching quorum and termination}
\ta{This phase relies on the \getThreshSig function described in \cref{algo:helpers}, which, given a commitment $C$, either returns a threshold signature for $C$ (received beforehand or aggregating $\tau=\lfloor\frac{n+t}{2}\rfloor+1$ signature shares stored for $C$) if it exists, or $\bot$ otherwise.}
This phase focuses on ensuring that, once a Byzantine quorum \ta{(represented by the threshold signature returned by \getThreshSig)} and enough message fragments for reconstructing the original message $m$ are gathered, at least~$\ell$ correct nodes deliver~$m$ and terminate.
Node~$p_i$ enters Phase~II only when there is a \ta{commitment $C'$ for which \getThreshSig returns a valid threshold signature}, and $p_i$ stores at least $k$~message fragments.
As long as no application message from~\psender was delivered (\cref{line:quorum:sigs:enough:fragments}), $p_i$~reconstructs the application message~$m_i$ (\cref{line:v3.1:reconstruct:m}) using the stored message fragments, 
and use this message as an input to \ta{\computeFragCom} (\cref{line:v3.1:recomputing:frags:with:proofs}), which outputs its \ta{commitment $C=\Commitment(m_i)$} along with coded message fragments and proofs of inclusion, $(\tilde m'_j, \pi'_j, j)$.
Node $p_i$ then ensures that the computed \ta{commitment $C$} matches the stored \ta{commitment $C'$} (\cref{line:hashes:equal:b4:delivery}). 
If this condition holds true, then $m_i=m$ is the message sent by the sender\footnote{To see why this is needed, consider a Byzantine sender that disseminates fragments $\tilde m_1,\ldots,\tilde m_n$, that do not form a proper codeword.}, and in particular, $p_i$ now holds \emph{all} the possible fragments for~$m$ along with their valid proof of inclusion, including fragments it has never received before!
Node~$p_i$ then \ta{retrieves the threshold signature $\Sigma_C$ of $C$ using the \getThreshSig function (\cref{line:get-tsig}), and disseminates it along with the message fragments} to the rest of the network. 
In particular, to each~$p_j$ in the network, $p_i$ sends a \bundle message (\cref{line:v3.1:bundle:comm})  that includes the \ta{commitment $C$}, fragment details $(\tilde m'_i, \pi'_i, i)$ and $(\tilde m'_j, \pi'_j, j)$, and the \ta{associated threshold signature~$\Sigma_C$}. 
After these transmissions, $p_i$ can MBRB-deliver the reconstructed message $m_i$ (\cref{ln:MBRBdeliverMjSn}).


The parameter $\ECCthreshold$ used at \cref{line:quorum:sigs:enough:fragments} is the number of (valid) fragments sufficient to reconstruct the application message~$m$ by the error-correction code~$\ECC$. 
This parameter should be practically selected by the desired $\ell$ given in \cref{thm:Global}. 
That is, one needs to set $\eps>0$ for $\ell = n-t-(1+\eps)\Dadv$ and then choose $k \le 1+ \frac{\eps}{1+\eps}(n-t-\Dadv)$. See details in \cref{sec:select-k}.

Upon the arrival of a \bundle message, \ie $\langle \bundle, \ta{C'}, (\tilde m'_j,\pi'_j,j), \frag'_i, \ta{\Sigma} \rangle$ arriving from $p_j$ (\cref{line:arrival:BUNDLE}), the recipient~$p_i$ validates the received message using the \isValid function (\ta{with the \isThreshSig parameter set to \true to indicate that we verify a threshold signature}) and discards invalid messages (\cref{line:isValid:in:BUNDLE}). 
%
Node $p_i$ proceeds to store the arriving information regarding the message segments $(\tilde m'_j,\pi'_j,j)$ and \ta{threshold signature $\Sigma$} for the specific \ta{commitment $C'$} (\cref{line:store:in:BUNDLE}). 
In the case that no \bundle message was sent by $p_i$ and the received $\frag'_i$ is nonempty (so $p_i$ learns its fragment, which it stores at \cref{line:store:my:fragment:in:BUNDLE}, unless already known), $p_i$ broadcasts a $\langle \bundle, \ta{C'}, (\tilde m'_i,\pi'_i,i), \bot, \ta{\Sigma} \rangle$ message (\cref{line:v3.1:bundle:with:bot}). 


The use of $\bot$ values appears also in \bundle messages (\cref{line:v3.1:bundle:comm,line:v3.1:bundle:with:bot}). 
A \bundle message might contain up to two fragments: the sender's fragment ($\tilde m'_i$ in the pseudo-code), which is always included, and the receiver's fragment ($\tilde m'_j$), which is included only when the sender was able to reconstruct the message $m$ (at \cref{line:v3.1:reconstruct:m}). 
The sender's fragments are collected by the receivers and allow reconstruction of the message once enough \bundle messages are received. 
The receiver's fragment allows the receiver to send \bundle messages (with its fragment), facilitating the dissemination of both \ta{threshold signatures} and fragments.

\Subsection{The error-correction code in use}
\ta{\computeFragCom (\cref{algo:helpers})} uses an error-correction code at \cref{ln:tildeMgetsECCm} to encode the application message~$m$, before it is split into $n$~fragments 
that will be disseminated by the parties. 
\ft{The code uses a fixed} parameter $\ECCthreshold$, that can be set later. Our algorithm requires that the ECC will be able to decode the message~$m$ from any subset of~$k$ fragments out of the $n$ fragments generated at \cref{ln:letM1ldots:new}. That is, we need an ECC that can deal with erasures, where the erased symbols are those contained in the $n-k$ missing fragments.
To that end, \ft{we use} a Reed-Solomon code $\ECC : \F^{\kECC} \to \F^{\nECC}$ with $\kECC > |m|/\log |\F|$. Each fragment contains $\nECC/n$ symbols of the codeword, and to be able to recover from $(n-k)\cdot \nECC/n$ erased symbols by   \cref{fct:ErasureECC}, we can set the code's distance to be $\dECC > (n-k)\cdot \nECC/n$. Since a Reed-Solomon code is MDS (see \cref{sec:ecc}),
$\dECC \le \nECC-\kECC+1$, and we can set $\nECC > \frac{n}{k}(\kECC-1)$. 
The code will have a constant rate, \ie $|\ECC(m)| = \bigO{|m|}$, as long as 
$m$ is sufficiently long, \ie $|m|=\Omega$($n\log|\F|$),  which implies that $\kECC=\Omega$($n$), and as long as $k=\Omega$($n$).
Recall also that $|\F|\ge \nECC$ is in a Reed-Solomon code.




	\newcommand{\numberBundleMsg}{\#\bundle}
	\newcommand{\bundleReceiveSet}{\ensuremath{B_{\mathsf{recv}}}\xspace}
	\newcommand{\bundleKReceiveSet}{\ensuremath{B_{\ECCthreshold,\mathsf{recv}}}\xspace}
	\newcommand{\bundleSendSet}{\ensuremath{B_{\mathsf{send}}}\xspace}

\Subsection{Analysis}
\label{sec:complexity}
The following main theorem states that our algorithm is correct.

\compact{\B}
\begin{theorem}[main]
	\label{thm:main-inf}
	Assume $n>3t+2d$,  $k\le (n-t-2d)$ and $\eps>0$.  	\Cref{alg:3.1:merkle:trees} implements MBRB with $\ell>n-t-(1+\eps)\Dadv$. Any algorithm activation on the input message~$m$ communicates $4n^2$ messages, where each node communicates \ta{$\bigO{|m|+n\kappa}$} bits.
\end{theorem}
\compact{\B}
For the sake of presentation, the complete proof and a discussion on the assumptions taken appear in Appendix~\ref{sec:correctness}. 
Here, we sketch the proof of MBRB-Global-delivery property in \cref{thm:SGlobal} (assuming the other properties hold)  and analyze the communication costs in \cref{lem:communication}.

\B
\begin{lemma}[MBRB-Global-delivery]
	\label{thm:SGlobal}
	Suppose a correct node, $p_i$, MBRB-delivers an application message~$m$ (\cref{ln:MBRBdeliverMjSn}).
	At least $\ell=\displaystyle c - \Dadv/(1-((\ECCthreshold-1)/(c-\Dadv)))$ correct nodes MBRB-deliver~$m$.
\end{lemma}
\B

\renewcommand{\lemcnt}{\ref{thm:SGlobal}}
\begin{lemmaProofSketch}
\sloppy
	Let \ta{$C_m=\Commitment(m)$}.
		The proof counts the \bundle messages disseminated by correct nodes. 
		If a correct node disseminates a \bundle message both at \cref{line:v3.1:bundle:comm,line:v3.1:bundle:with:bot}, we only consider the one from \cref{line:v3.1:bundle:comm}.
		Let \bundleSendSet be the set of correct nodes that disseminate at least one \bundle message, let \bundleReceiveSet be the set of correct nodes that receive at least one valid \bundle message from a correct node during their execution, and let \bundleKReceiveSet be the set of correct nodes that receive \bundle messages from at least $\ECCthreshold$ distinct correct nodes.
		The following holds.	
		
\B		\begin{subobservation}\label{subobs:SbundleR:and:bundleS:geq:c-d}
			$c \geq |\bundleReceiveSet| \geq c-\Dadv$ and $c \geq |\bundleSendSet| \geq c-\Dadv$.
		\end{subobservation}
		

		\renewcommand{\obscnt}{\ref{subobs:SbundleR:and:bundleS:geq:c-d}}
\B		\begin{obsProof}
			Since \bundleSendSet and \bundleReceiveSet contain only correct nodes, trivially $c \geq |\bundleSendSet|$ and $c \geq |\bundleReceiveSet|$.
            \ft{$|\bundleReceiveSet| \geq c-\Dadv$ and $|\bundleSendSet| \geq c-\Dadv$ follow from the definition of the message adversary, and the way the algorithm chains \bundle messages
            .}
		\end{obsProof}
\B

\B		
		\begin{subobservation}\label{subobs:Scounting:bundle:equation}
			$|\bundleKReceiveSet|\times|\bundleSendSet| + (\ECCthreshold-1)(|\bundleReceiveSet|-|\bundleKReceiveSet|) \geq |\bundleSendSet|(c-\Dadv)$.
		\end{subobservation}
\B		

	\renewcommand{\obscnt}{\ref{subobs:Scounting:bundle:equation}}
\B	\begin{obsProof}
        \ft{The inequality follows from a counting argument on the overall number of valid \bundle messages received by correct nodes from distinct correct senders. 
        In particular, we use the fact that nodes in $\bundleReceiveSet\setminus\bundleKReceiveSet$ each receives at most $\ECCthreshold-1$ valid \bundle messages from distinct correct senders.}
	\end{obsProof}
\B

\BB	
	\begin{subobservation}\label{subobs:SbundleKReceiveSet:geq:frac}
		$\displaystyle |\bundleKReceiveSet| \geq c - \Dadv/(1-((\ECCthreshold-1)/(c-\Dadv))).$
	\end{subobservation}
\B	

	\renewcommand{\obscnt}{\ref{subobs:SbundleKReceiveSet:geq:frac}}
	\begin{obsProof}
        \ft{The observation is obtained by isolating $\bundleKReceiveSet$ in \cref{subobs:Scounting:bundle:equation}, and minimizing the right-hand side to remove the dependence on $|\bundleSendSet|$.}
%
%
%
	\end{obsProof}
\B
\BB	\begin{subobservation}\label{subobs:Sall:bundleKReceiveSet:mbrb:deliv:m}
		All nodes in $\bundleKReceiveSet$ MBRB-deliver~$m$.
	\end{subobservation}
\B	
	\renewcommand{\obscnt}{\ref{subobs:Sall:bundleKReceiveSet:mbrb:deliv:m}}
	\begin{obsProof}
        \ft{The observation results from the properties of the vector commitment scheme, the unforgeability of signatures, and the ability of the ECC scheme to reconstruct $m$ from $\ECCthreshold$ distinct valid fragments for $m$.}
	\end{obsProof}
%

As all nodes in $\bundleKReceiveSet$ are correct, the above observations  yield the lemma.
\end{lemmaProofSketch}	

\remove{ 

Let $p_i$ be a correct node that MBRB-delivers $m$ (\cref{ln:MBRBdeliverMjSn}).  
	Let $h_m$ be the Merkle root hash returned by $\computeFragmentMerkleTree()$ for $m$.
	We count the number of \bundle messages disseminated by correct nodes (\cref{line:v3.1:bundle:comm,line:v3.1:bundle:with:bot}). 
 %
%
%
	Let \bundleSendSet be the set of correct nodes that disseminate at least one \bundle message during their execution. Similarly, let \bundleReceiveSet be the set of correct nodes that receive at least one valid \bundle message from a correct node during their execution. The detailed proof shows $c \geq |\bundleReceiveSet| \geq c-\Dadv$ and $c \geq |\bundleSendSet| \geq c-\Dadv$.
	%
	Let \bundleKReceiveSet be the set of correct nodes that receive \bundle messages from at least $\ECCthreshold$ distinct correct nodes.
	The detailed proof shows $|\bundleKReceiveSet|\times|\bundleSendSet| + (\ECCthreshold-1)(|\bundleReceiveSet|-|\bundleKReceiveSet|) \geq |\bundleSendSet|(c-\Dadv)$.
%
%
\begin{wrapfigure}{r}{0.475\textwidth}
		\centering
		\B\compact{\BBB}
		\begin{footnotesize}
				\begin{tikzpicture}[scale=0.5]
    \tikzmath{
    \heightRcvk = 2.5;
    \heightB = 1.25;
    \widthRcvk = 2.5;
    \widthRcv = 5;
    }
    
    \def\colorA{black}
    \def\colorB{black}

    \draw[-{>[scale=2.5,length=2,width=3]}] (0,-.25) -- (0,\heightRcvk+.5);
    \node[align=center] at (-3.5,\heightRcvk) {\# received\\\bundle};
    \draw[-{>[scale=2.5,length=2,width=3]}] (-.25,0) -- (\widthRcv+1,0);
    \node[align=center] at (\widthRcv+2,.3) {\# correct\\nodes};
    \draw (\widthRcv,0) -- (\widthRcv,-.1);
    
    \node at (-1.1,\heightRcvk) {$|\bundleSendSet|$};
    \draw (-.1,\heightRcvk) -- (\widthRcvk,\heightRcvk) -- (\widthRcvk,-.1);
    \node at (\widthRcvk,-.4) {$|\bundleKReceiveSet|$};
    
    \node at (-.7,\heightB) {\ECCthreshold-1};
    \draw (0,\heightB) -- (-.1,\heightB);
    \draw[dashed] (0,\heightB) -- (\widthRcvk,\heightB);
    \draw (\widthRcvk,\heightB) -- (\widthRcv,\heightB) -- (\widthRcv,-.1);
    \node at (\widthRcv,-.4) {$|\bundleReceiveSet|$};

    \draw (0,-1.5) -- (0,-1.5);
\end{tikzpicture}
		\end{footnotesize}
	\BBB\compact{\BBB}
	\caption{\smaller{\bundle messages received by correct nodes.}}
		\label{fig:Smsg_dist}
  \compact{\BBB}\BB
	\end{wrapfigure}
%
		To see why the latter is true, let \numberBundleMsg be the number of valid \bundle messages received by correct nodes from distinct correct senders.
		As the message adversary may drop up to~$\Dadv$ out of the $n$ messages of this $\comm$, we are guaranteed that at least $c-\Dadv$ correct nodes receive $p$'s \bundle message. This immediately implies that $\numberBundleMsg \geq |\bundleSendSet|(c-\Dadv)$.
%
%
		As illustrated in \cref{fig:Smsg_dist}, the nodes in $\bundleKReceiveSet$ may receive up to $|\bundleSendSet|$ valid \bundle messages from distinct correct senders
  %
  %
  for a maximum of $|\bundleKReceiveSet|\times|\bundleSendSet|$ \bundle messages overall. The remaining node of $\bundleReceiveSet\setminus\bundleKReceiveSet$ may each receive up to $\ECCthreshold-1$ valid \bundle messages from distinct correct senders, by definition of $\bundleKReceiveSet$. 
		As $\bundleKReceiveSet\subseteq\bundleReceiveSet$ by definition, $|\bundleReceiveSet\setminus\bundleKReceiveSet|=|\bundleReceiveSet|-|\bundleKReceiveSet|$, and the nodes of $\bundleReceiveSet\setminus\bundleKReceiveSet$
		accounts for up to $(\ECCthreshold-1)(|\bundleReceiveSet|-|\bundleKReceiveSet|)$ valid \bundle messages overall. As the \bundle messages counted by \numberBundleMsg are received either by correct nodes in $\bundleKReceiveSet$ or in $\bundleKReceiveSet\setminus\bundleReceiveSet$, these observations lead to
			$|\bundleKReceiveSet|\times|\bundleSendSet| + (\ECCthreshold-1)(|\bundleReceiveSet|-|\bundleKReceiveSet|) \geq \numberBundleMsg$.
		

		The rest of the proof follows from the observations that 
  $\displaystyle |\bundleKReceiveSet| \geq c - \Dadv/{(1-((\ECCthreshold-1)/(c-\Dadv)))}$,
%
%
  all nodes in $\bundleKReceiveSet$ MBRB-deliver~$m$, and the fact that all nodes in $\bundleKReceiveSet$ are correct.
This completes the sketch of the proof of MBRB-Global-delivery.

} 

\smallskip

\B

\begin{lemma}\label{lem:communication}
	Correct nodes collectively communicate at most~$4n^2$ messages. Each correct node \ta{sends at most $\bigO{|m| + n\kappa}$} bits. Overall, \ta{the system sends at most $\bigO{n|m| + n^2\kappa}$} bits. 
\end{lemma}
\compact{\B}
\renewcommand{\lemcnt}{\ref{lem:communication}}
\begin{lemmaProof}
	Let us count the messages communicated by counting $\comm$ and $\broadcast$ invocations.
	The sender,~\psender, sends \send messages at \cref{ln:v3.1:commV1ldotsVn}.
	In Phase~I, each correct node that has received a \send message broadcasts a \forward message once (\cref{line:v3.1:bcast:forward,line:condition:sign:hash:SEND}). 
	However, if it receives a \forward before the \send arrives, it performs one additional \forward broadcast (\cref{line:v3.1:forward:after:forward}). 
	This yields at most $2$ $\comm$ and $\broadcast$ invocations per correct node until the end of Phase~I. 
	We can safely assume that a correct sender always sends a single \forward (\ie it immediately and internally receives the \send message sent to self).
	Thus, $\psender$ is also limited to at most $2$ invocations up to this point.
	In Phase~II, each correct node that MBRB-delivers a message at \cref{ln:MBRBdeliverMjSn} transmits \bundle messages at \cref{line:v3.1:bundle:comm}.
	This can only happen once due to the condition at \cref{ln:bulljsnWasNotMBRBdelivered}.
	Additionally, it may transmit \bundle messages also at \cref{line:v3.1:bundle:with:bot}, upon the reception of a \bundle. 
	However, this second \bundle transmission can happen at most once, due to the if-statement at \cref{line:test:if:already:bundle}.
	This leads to at most $2$ additional $\comm$ and $\broadcast$ invocations per correct node.
	Thus, as the number of correct nodes is bounded by~$n$, the two phases incur in total at most $4n$ invocations of $\comm$ and $\broadcast$ overall. 
	Since each invocation communicates exactly $n$ messages, the total message cost
	for correct nodes when executing one instance of \cref{alg:3.1:merkle:trees} is upper bounded by~$4n^2$.
	Note that the above analysis holds for correct nodes also in the presence of Byzantine participants, including when \psender is dishonest.
	
	We now bound the number of bits communicated by correct nodes throughout a single instance of \cref{alg:3.1:merkle:trees}.
	Consider \ta{\computeFragCom}. 
	Let $m$ be a specific application message. 
	We have $|\tilde m|=\bigO{|m|}$ since we use a code with a constant rate. 
	Thus, any specific fragment $\tilde m_i$ has length $|\tilde m_i| = \bigO{|m|/n}$.
	Recall that \ta{the sizes of a signature share $\sigma$, a threshold signature $\Sigma$, a commitment $C$, and an inclusion proof $\pi$ all have} $\bigO{\kappa}$~bits (\cref{sec:crypto-prim}).
	Along \ta{a signature share pair $\sig$}, the identifier of the signing node is included, which takes additional $\bigO{\log n}$~bits. 
	However, since $\kappa=\Omega$($\log n$), the inclusion of this field does not affect asymptotic communication costs.
	
	We now trace all the $\comm$ and $\broadcast$ instances in \cref{alg:3.1:merkle:trees} and analyze the number of bits communicated in each.
	The \send $\comm$ (\cref{ln:v3.1:commV1ldotsVn}) communicates $n$ messages, where each message includes a fragment of $m$ ($\bigO{|m|/n}$ bits) with its \ta{proof of inclusion ($\bigO{\kappa}$ bits)}, a \ta{commitment} ($\bigO{\kappa}$ bits), and a \ta{signature share} ($\bigO{\kappa}$ bits).
    Thus, this operation allows the sender to communicate at most \ta{$\bigO{|m|+ n\kappa}$} bits. 
	Each \forward broadcast in lines~\ref{line:v3.1:bcast:forward} and~\ref{line:v3.1:forward:after:forward} sends $n$ copies of a message containing a \ta{commitment} ($\bigO{\kappa}$ bits), at most one message fragment with its \ta{proof of inclusion ($\bigO{|m|/n+\kappa}$ bits)}, and two \ta{signature shares} ($\bigO{\kappa}$ bits).
	Hence, each one of lines~\ref{line:v3.1:bcast:forward} and~\ref{line:v3.1:forward:after:forward} communicates a total of \ta{$\bigO{|m| + \kappa n}$} bits.
	The \bundle communication (lines~\ref{line:v3.1:bundle:comm} or~\ref{line:v3.1:bundle:with:bot}) sends $n$ messages, where each contains a \ta{commitment} ($\bigO{\kappa}$ bits), at most two message fragments with their \ta{proof of inclusion ($\bigO{|m|/n+ \kappa}$ bits)}, and \ta{one threshold signature ($\bigO{\kappa}$ bits)}.
    Hence, each line communicates at most \ta{$\bigO{|m|+n\kappa}$} bits.
	As analyzed above, the sending node (\psender, when correct) 
	performs at most one $\comm$ of \send messages, while each correct node performs at most two $\broadcast$ of \forward messages, and at most two $\comm$/$\broadcast$ of \bundle messages. 
	Thus, each node communicates at most \ta{$\bigO{|m| + n\kappa}$} bits.
	Overall, the total bit communication by correct nodes during \cref{alg:3.1:merkle:trees}'s execution is \ta{$\bigO{n|m| + n^2\kappa}$}.
	As mentioned above, the analysis holds in the presence of Byzantine nodes, even if \psender is dishonest.
\end{lemmaProof}

\Section{Conclusion}
We introduced a Coded MBRB algorithm that significantly improves the state-of-the-art solution. 
It achieves \ta{optimal communication (up to the size of cryptographic parameter~$\kappa$)} while maintaining a high delivery power, \ie it ensures that messages are delivered by at least $\ell = n-t-(1+\eps)d$ correct nodes, where $\eps>0$ is a tunable parameter.
The proposed solution is deterministic up to its use of \ta{cryptography (threshold signatures and vector commitments)}.
Each correct node sends no more than $4n$ messages and communicates at most \ta{$\bigO{|m|+n\kappa}$} bits, where $|m|$ represents the length of the input message and $\kappa$ is a security parameter. 
We note that the algorithm's communication efficiency holds for sufficiently long messages and approaches the natural upper bound on delivery power, $ n-t-Dadv$, which accounts for the message adversary's ability to isolate a subset of correct nodes.
The proposed approach achieves a delivery power $\ell$ that can be made arbitrarily close to this limit, albeit with a marginal increase in communication costs, which depends on the chosen  $\eps$. This work represents a significant advancement in Byzantine Reliable Broadcast, offering a practical solution for robust communication in asynchronous message-passing systems with malicious nodes and message adversaries.
One intriguing question is whether \ta{it is possible to devise an (M)BRB algorithm that does not exhibit the $\kappa$ parameter in its communication complexity or the $\epsilon$ parameter in its delivery power $\ell$, for instance by leveraging randomization~\cite{DBLP:journals/dc/AbrahamCDNPRS23} or error-freedom~\cite{DBLP:conf/podc/AlhaddadDD0VXZ22}.}
\newpage

\bibliographystyle{plain}
\bibliography{arXiv}


\newpage

\appendix

\section{Correctness analysis}
\label{sec:correctness}
We now analyze \cref{alg:3.1:merkle:trees} and show it satisfies the MBRB properties specified in
\cref{def:mbrb}. 
This analysis, along with the communication analysis in \cref{lem:communication} prove our main Theorem~\ref{thm:main-inf}.
\begin{assumption}[coded-MBRB-assumption]
	\label{asm:MBRBassumptionR}
	$n>3t+2\Dadv$ and $\ECCthreshold\leq (n-t)-2\Dadv$.
\end{assumption}	
\begin{theorem}
	\label{thm:main}
	For any network that satisfies \cref{asm:MBRBassumptionR} and for any $\eps>0$, 
	\Cref{alg:3.1:merkle:trees} implements an MBRB algorithm (\cref{def:mbrb}) with $\ell>n-t-(1+\eps)\Dadv$. 
\end{theorem}

Recall that any MBRB algorithm requires $n>3t+2\Dadv$ \cite{AFRT23}. Our coded MBRB algorithm uses an error-correction code that can reconstruct any encoded message from $k$ fragments of the codeword. 
While we have some flexibility in selecting the value of~$k$, which affects the parameters of the ECC and thus the communication complexity, our proof requires that $k$ will not be too large. 
We begin with a few technical lemmas.

\begin{lemma}\label{lemma:stored:fragments:are:valid}
	If a correct node $p_u$ stores a message fragment $\tilde m_j$ associated to a proof of inclusion $\pi_j$ for some \ta{commitment $C'$} and node identity $j$, then $\pi_j$ is valid with respect to~\ta{$C'$}, that is $\ta{\vcVerify(C',\pi_j,\tilde m_j,j)}=\valid$.
\end{lemma}
\renewcommand{\lemcnt}{\ref{lemma:stored:fragments:are:valid}}
\begin{lemmaProof}
    A correct node stores fragments for a \ta{commitment $C'$} at \cref{line:store:in:SEND,line:store:my:fragment:in:FORWARD,line:store:in:BUNDLE,line:store:my:fragment:in:BUNDLE}, when receiving \send, \forward, or \bundle messages, respectively.
    The fragments stored at these lines and their proof have been received through the corresponding message, whose content is verified by a call to \isValid (at \cref{line:isValid:in:SEND,line:isValid:in:FORWARD,line:isValid:in:BUNDLE}). \isValid (described in \cref{algo:helpers}) checks that proofs of inclusion are valid for the corresponding \ta{commitment}.
\end{lemmaProof}

The following notion of \emph{valid messages} will be used throughout the analysis to indicate messages containing only valid information, as the algorithm dictates.
\begin{definition}[Valid messages]\label{def:valid}
	We say that a message of type \send, \forward, or \bundle is \emph{valid} if and only if \isValid returns $\true$ at line~\ref{line:isValid:in:SEND}, \ref{line:isValid:in:FORWARD}, or \ref{line:isValid:in:BUNDLE}, respectively, upon the receipt of that message.
\end{definition}
Operatively, valid messages satisfy the following which is immediate from the definition of the \isValid function (\cref{algo:helpers}).
\begin{corollary}\label{cor:valid}
	To be valid, a message must meet the following criteria: (i) all the \ta{signatures shares or threshold signatures} it contains must be valid and correspond to the \ta{commitment} included in the message; (ii) \ta{if it is of type \send or \forward,} it must contain a signature by the designated sending node \psender; and (iii) all inclusion proofs must be valid with respect to the \ta{commitment} included in the message.
\end{corollary}

We now show that the correct parties always send valid messages. However, they might receive invalid messages sent by Byzantine nodes. 
\begin{lemma}\label{lem:all:msgs:sent:valid}
	All \send, \forward, or \bundle messages sent by a correct node~$p_u$, are valid.  
\end{lemma}
\renewcommand{\lemcnt}{\ref{lem:all:msgs:sent:valid}}
\begin{lemmaProof}
	The only correct node that sends \send messages is~$\psender$ at \cref{ln:v3.1:sendFrag}. Indeed, \ft{when $\psender$ is correct,} this message will contain a valid \ta{signature share} by~$\psender$ and all \ta{proofs of inclusion} are valid, by \cref{line:fragments:for:m,line:ps:signs:hash}.

	Now consider a \forward message sent either at lines~\ref{line:v3.1:bcast:forward} or~\ref{line:v3.1:forward:after:forward}.
    To reach there, $p_u$ \ft{must have} passed \cref{line:isValid:in:SEND} or \cref{line:isValid:in:FORWARD}, which guarantees $p_u$ received a valid signature for~\ta{$C'$} made by~$\psender$ (where \ta{$C'$} is the \ta{commitment} in the received message triggering this code). 
	Then, at \cref{line:store:in:SEND} or at \cref{line:store:in:FORWARD}, $p_u$ stores a signature of~$\psender$ for this~\ta{$C'$}, and at \cref{line:sign:a:hash:in:SEND,line:sign:a:hash:in:FORWARD}, $p_u$  signs the same~\ta{$C'$}.
    Thus, conditions (i) and (ii) of \cref{cor:valid} hold, and if the \forward is sent at \cref{line:v3.1:forward:after:forward}, then condition (iii) vacuously holds as well.
	If the \forward message is sent at \cref{line:v3.1:forward:after:send}, it contains a fragment that was stored by~$p_u$ for the same~\ta{$C'$}, and by \cref{lemma:stored:fragments:are:valid}, its associated proof of inclusion is valid; thus condition (iii) holds in this case as well.
	
    Finally, consider a \bundle message.
    \ta{
    First off, this type of message is not concerned by condition~(ii) of \cref{cor:valid}.
    For the transmission at \cref{line:v3.1:bundle:comm}, condition (i) follows from the construction of the threshold signature $\Sigma_C$ at \cref{line:get-tsig}. \ft{$\Sigma_C$ is guaranteed to be non-$\bot$ by the condition at \cref{line:quorum:sigs:enough:fragments} of \cref{alg:3.1:merkle:trees}, and is provided by the helper function $\getThreshSig(\cdot)$ (\cref{line:return:TS:from:storage} of \cref{algo:helpers}).}
    \ft{When executing $\getThreshSig(\cdot)$,} the first possibility is that $\Sigma_C$ \ft{is already known by $p_u$ because it} was received by $p_u$ at \cref{line:arrival:BUNDLE} and stored at \cref{line:store:in:BUNDLE}.
    In this case, the validity of $\Sigma_C$ is ensured by the check at \cref{line:isValid:in:BUNDLE}.
    The second possibility is that $\Sigma_C$ aggregates $\tau=\lfloor\frac{n+t}{2}\rfloor+1$ signature shares received by $p_u$ at \cref{ln:sendRx:SEND} or \cref{line:arrival:FORWARD}, and stored at \cref{line:store:in:SEND} or \cref{line:store:in:FORWARD}, respectively.
    In this case, the validity of all these signature shares is ensured by the checks at \cref{line:isValid:in:SEND} and \cref{line:isValid:in:FORWARD}, respectively, and thus the aggregated threshold signature $\Sigma_C$ is also valid.
    }
    Condition (iii) follows since the \ta{proofs of inclusion} were computed at \cref{line:v3.1:recomputing:frags:with:proofs} by $p_u$ and match the same \ta{commitment~$C'$} used in that \bundle message, as enforced by \cref{line:hashes:equal:b4:delivery}. 
	\ta{For the broadcast at \cref{line:v3.1:bundle:with:bot}, conditions (i) and (iii) follow since the threshold signature $\Sigma$ and the fragment tuple $(\tilde m'_j,\pi'_j,j)$ come from the incoming \bundle message at \cref{line:arrival:BUNDLE}, whose validity (w.r.t. $C'$) has been verified at \cref{line:isValid:in:BUNDLE}.
    }
	%
\end{lemmaProof}

\begin{lemma}\label{lem:single:hash:sign}
	A correct node $p$ signs a most a single \ta{commitment~$C$}.
\end{lemma}
\renewcommand{\lemcnt}{\ref{lem:single:hash:sign}}
\begin{lemmaProof}
	$p$ signs a \ta{commitment} either at \cref{line:ps:signs:hash} (for \psender), \ref{line:sign:a:hash:in:SEND} or \ref{line:sign:a:hash:in:FORWARD}. We consider two cases, depending on whether $p$ is \psender or not.
	\begin{itemize}
		
		\item {\em Case 1:} Assume $p\neq\psender$. $p$ can  sign some \ta{commitment} only at \cref{line:sign:a:hash:in:SEND} or \ref{line:sign:a:hash:in:FORWARD}. By the conditions at \cref{line:condition:sign:hash:SEND,line:condition:sign:hash:FORWARD}, \cref{line:sign:a:hash:in:SEND,line:sign:a:hash:in:FORWARD} are executed only if either $p$ has not signed any \ta{commitment} yet, or has already signed the exact same \ta{commitment $C'$}. 
		
		\item {\em Case 2:} If $p=\psender$, because valid messages must contain $\psender$'s \ta{signature share} (due to calls to $\isValid()$ at \cref{line:isValid:in:SEND,line:isValid:in:FORWARD}), and because we have assumed that signatures cannot be forged,
		\cref{line:ps:signs:hash} is always executed before \cref{line:condition:sign:hash:SEND,line:v3.1:test:forward:sent}. By the same reasoning as Case 1, \psender therefore never signs a different \ta{commitment} at \cref{line:sign:a:hash:in:SEND} or \ref{line:sign:a:hash:in:FORWARD}. \qedhere 
	\end{itemize}%
\end{lemmaProof}

We recall that the above lemmas, and as a consequence, the upcoming theorems, hold with high probability, assuming a computationally-bounded adversary that forges \ta{signature shares/threshold signatures} or finds \ta{commitment} collisions with only negligible probability.
We can now prove the properties required for an MBRB algorithm, as depicted in \cref{def:mbrb}.
\begin{lemma}[MBRB-Validity]
	\label{thm:Validity}
	Suppose \psender is correct and a correct node, $p_i$, MBRB-delivers an {application message}, $m$. Then, node~\psender has previously MBRB-broadcast~$m$.
\end{lemma}
\renewcommand{\lemcnt}{\ref{thm:Validity}}
\begin{lemmaProof}
	Suppose $p_i$ MBRB-delivers $m$ at \cref{ln:MBRBdeliverMjSn}. \ta{Consider $C'$ the commitment} that renders true the condition at \cref{line:quorum:sigs:enough:fragments}, and \ta{$C$ the commitment} that is computed at \cref{line:v3.1:recomputing:frags:with:proofs}. It holds that \ta{$C'=C$} by \cref{line:hashes:equal:b4:delivery}, or otherwise $p_i$ could not have reached \cref{ln:MBRBdeliverMjSn}.
	%
	
	\ft{Consider} the \ta{threshold signature $\Sigma_C$ returned by the \getThreshSig function at \cref{line:get-tsig}. \ft{Using the same reasoning as in the proof of \cref{lem:all:msgs:sent:valid}, $\Sigma_C$} is valid, \ft{and} must\ft{, therefore,} aggregate at least $\tau=\lfloor\frac{n+t}{2}\rfloor+1$ valid signature shares for $C$.
    Let us remark that, out of all these valid signature shares, at least $\lfloor\frac{n+t}{2}\rfloor+1-t = \lfloor\frac{n-t}{2}\rfloor+1 \geq 1$ are generated by correct nodes\footnote{
        Remind that, $\forall x \in \mathbb{R}, i \in \mathbb{Z}: \lfloor x \rfloor+i = \lfloor x+i \rfloor$.
    }.
    Thus, at least one correct node $p_j$ must have produced a signature share for $C$, whether it be at \cref{line:sign:a:hash:in:SEND} or \cref{line:sign:a:hash:in:FORWARD} if $p_j \neq \psender$, or at \cref{line:ps:signs:hash} if $p_j = \psender$.
    However, in all these cases, the sender \psender must have necessarily produced a signature share for $C$: the case $p_j=\psender$ is trivial, and in the case $p_j \neq \psender$, $p_j$ must have verified the presence of \ft{a valid signature share} from \psender in the message it received, at \cref{line:isValid:in:SEND} or \cref{line:isValid:in:FORWARD}, respectively.
    }
	
	Under the assumption that the adversary cannot forge \ta{signature shares/threshold signatures} (\cref{sec:signatures}), and recalling that \psender is correct, the only way in which \psender could have signed~\ta{$C'$} is by executing \cref{line:ps:signs:hash} when MBRB-broadcasting some message $m'$ at \cref{ln:MBRBbroadcastMsn}; see also the proof of \cref{lem:single:hash:sign}.
	Furthermore, recall that the \ta{commitment is collision-resistant (or binding, see \cref{sec:vc})}, meaning that except with negligible probability, the message $m'$ that \psender uses in \cref{ln:MBRBbroadcastMsn} satisfies $m'=m$, since it holds that \ta{$C'=\Commitment(m')=\Commitment(m)=C$}.
\end{lemmaProof}



\begin{lemma}[MBRB-No-duplication]
	\label{thm:duplication}
	A correct node~$p_i$ MBRB-delivers at most one {application message},~$m$.
\end{lemma}
\renewcommand{\lemcnt}{\ref{thm:duplication}}
\begin{lemmaProof}
	The condition at \cref{ln:bulljsnWasNotMBRBdelivered} directly implies the proof.
\end{lemmaProof}

\begin{lemma}[MBRB-No-duplicity]
	\label{thm:duplicity}
	No two different correct nodes MBRB-deliver different application messages.
\end{lemma}
\renewcommand{\lemcnt}{\ref{thm:duplicity}}
\begin{lemmaProof}
	Suppose, towards a contradiction, that $p_i$ MBRB-delivers $m$ and $p_j$ MBRB-delivers $m'\neq m$, where $p_i$ and $p_j$ are both correct nodes.
    Let us denote by \ta{$C$, resp. $C'$, the commitment returned by $\computeFragCom()$} for $m$, resp. for $m'$.
    As \ta{commitments} are assumed to be collision-resistant (\cref{sec:vc}), $m \neq m'$ implies \ta{$C \neq C'$}.
	
    By the condition at \cref{line:quorum:sigs:enough:fragments}, $p_i$ \ta{gets a threshold signature $\Sigma_i \neq \bot$ from the \getThreshSig function that aggregates a set $Q_i$ \ft{containing} $\tau=\lfloor\frac{n+t}{2}\rfloor+1$ valid signature shares for $C$.
    Similarly, $p_j$ gets a threshold signature $\Sigma_j$ aggregating a set~$Q_j$ of signature shares for~$C'$. 
    }
	%
	%
	We know that $|Q_i\cup Q_j|=|Q_i|+|Q_j|-|Q_i \cap Q_j|$.
        \ta{Moreover, we know that, $\forall x \in \mathbb{R}, k \in \mathbb{Z}: k = \lfloor x \rfloor+1 \implies k > x$, and hence we have $Q_i > \frac{n+t}{2} < Q_j$.}
	Thus, $|Q_i \cap Q_j|\geq |Q_i|+|Q_j|-n > 2 \frac{n+t}{2}-n=t$.
	In other words, $Q_i$ and $Q_j$ have at least one \emph{correct} node, $p_u$, in common that has signed both \ta{$C$ and $C'$}. 
	\cref{lem:single:hash:sign}, and the fact that $p_u$ has signed both \ta{$C$ and $C'$} leads the proof to the needed contradiction.
    Thus, $m=m'$, and the lemma holds.
\end{lemmaProof}

\begin{lemma}[MBRB-Local-delivery]
	\label{thm:Local}
	\sloppy
	Suppose \psender is correct and MBRB-broadcasts $m$.
	At least one correct node, say, $p_j$, MBRB-delivers $m$.
\end{lemma}
\renewcommand{\lemcnt}{\ref{thm:Local}}
\begin{lemmaProof}
	Let us denote by \ta{$C_m$ the commitment} computed at \cref{line:v3.1:MHash} when executing \ta{$\computeFragCom(m)$}.
	%
	The proof of the lemma will follow from 
	\crefrange{subobs:all:msgs:contain:h:sig_s}{subobs:one:correct:delivers:m} stated and proven below.
	
	
	\begin{subobservation}\label{subobs:all:msgs:contain:h:sig_s}
		All valid \send, \forward, or \bundle messages received by some correct node $p_u$ contain~\ta{$C_m$}.
	\end{subobservation}
	
	\renewcommand{\obscnt}{\ref{subobs:all:msgs:contain:h:sig_s}}

\sloppy\begin{obsProof}
Recall that $\psender$ MBRB-broadcasts $m$, thus we know that $\psender$ has included its own \ta{signature share, $\sig_s=\big(\TSIGN_s(C_m),s\big)$}, when it propagates $\langle \send, \ta{C_m}, (\tilde m_j, \pi_j,j), \sig_s \rangle$ (\crefrange{line:fragments:for:m}{ln:v3.1:commV1ldotsVn}).
Consider a correct node $p_u$ that receives a valid \send, \forward, or \bundle message containing a \ta{commitment $C_u$} at lines~\ref{ln:sendRx:SEND}, \ref{line:arrival:FORWARD}, or~\ref{line:arrival:BUNDLE}. 
\ta{If the message is of type \send or \forward, then, as it is valid, it must contain $\psender$'s signature on~$C_u$.
If the message is of type \bundle, then, similarly to \cref{thm:Validity}, its valid threshold signature for $C_u$ aggregates a set of valid signature shares for $C_u$ that contains at least one share produced by a correct node $p_x$.
But for $p_x$ to produce this share, \psender must also have produced a valid signature share for $C_u$, either because $p_x$ must have checked its existence at \cref{line:isValid:in:SEND} or \cref{line:isValid:in:FORWARD} (prior to signing, at \cref{line:sign:a:hash:in:SEND} or \cref{line:sign:a:hash:in:FORWARD}, respectively), or because $p_x$ is the sender.
Hence, in any case, \psender produces a signature share for $C_u$.}
Since $\psender$ is correct, by \cref{lem:single:hash:sign}, it does not sign another \ta{commitment $C' \neq C_m$}. 
Under the assumption that signatures cannot be forged, the above implies that \ta{$C_u=C_m$}.
\end{obsProof}

	\begin{subobservation}\label{subobs:n:correct:sign:h':neq:h}
		A correct node~$p_u$ only signs \ta{valid signature shares} for~\ta{$C_m$}.
	\end{subobservation}
	
\renewcommand{\obscnt}{\ref{subobs:n:correct:sign:h':neq:h}}
\begin{obsProof}
    If $p_u=\psender$, it MBRB-broadcasts a single message and executes \cref{line:ps:signs:hash} only once, signing \ta{$C_m$}.
    Besides \cref{line:ps:signs:hash}, a correct node $p_u$ only signs \ta{signature shares} after receiving a valid \send or \forward message (at \cref{line:sign:a:hash:in:SEND,line:sign:a:hash:in:FORWARD}), and when it does, $p_u$ only ever signs the \ta{commitment} received in the message.
    By \cref{subobs:all:msgs:contain:h:sig_s}, this implies $p_u$ never signs any \ta{$C' \neq C_m$}.
\end{obsProof}
	
	
	\begin{subobservation}\label{subobs:all:correct:send:forward:form}
		If a correct node $p_u$ broadcasts a \forward message, this message is of the form $\langle \forward, \ta{C_m}, \bull, \{\sig_s,\sig_u\} \rangle$, where $\sig_s,\sig_u$ are $\psender$'s and $p_u$'s   valid \ta{signature shares} for~\ta{$C_m$}. 
	\end{subobservation}
	
	%
	
	\renewcommand{\obscnt}{\ref{subobs:all:correct:send:forward:form}}
	\begin{obsProof}
		Consider a correct node $p_u$ that broadcasts a message $\langle \forward,\allowbreak \ta{C'},\bull,\sigsset \rangle$ either at lines~\ref{line:v3.1:bcast:forward} or~\ref{line:v3.1:forward:after:forward}.
		By \cref{subobs:all:msgs:contain:h:sig_s}, \ta{$C'=C_m$}.
		The observation then follows from \cref{lem:all:msgs:sent:valid}.
	\end{obsProof}

	\newcommand{\frwdReceiveSet}{F_{\mathsf{recv}}}
	
	We now define $\frwdReceiveSet$ to be the set of correct nodes that receive a valid message $\langle \forward,\allowbreak \ta{C_m},  \bull, \sigsset \rangle$ at \cref{line:arrival:FORWARD}, where $\sigsset$ contains $\psender$'s valid signature for~\ta{$C_m$}. We analyze its size and the behavior of such nodes in the following observations.
	
	\begin{subobservation}\label{subobs:frwdReceiveSet:not:empty}
		$\frwdReceiveSet$ contains at least one correct node, \ie $\frwdReceiveSet\neq\emptyset$.
	\end{subobservation}
	\renewcommand{\obscnt}{\ref{subobs:frwdReceiveSet:not:empty}}
	\begin{obsProof}
		If \psender is correct and MBRB-broadcasts $m$, it executes \cref{ln:v3.1:commV1ldotsVn} and disseminates messages of the form $\langle \send, \ta{C_m}, (\tilde m_j, \pi_j), \sig_s \rangle$ to all nodes, where $\sig_s$ is \psender's \ta{signature share of $C_m$}.
        By definition of the message adversary, these \send messages are received by at least $c-d$ correct nodes. 
		
		By \cref{asm:MBRBassumptionR}, $n>3t + 2{\Dadv}$, and therefore $c-d\geq n-t-d > 0$. At least one correct node $p_x$, therefore, receives one of the \send messages disseminated by \psender at \cref{ln:v3.1:commV1ldotsVn}.
        As \psender is correct, by \cref{lem:all:msgs:sent:valid}, this message is valid, and is handled by $p_x$ at \crefrange{ln:sendRx:SEND}{line:v3.1:forward:after:send}. 
		
		By \cref{subobs:n:correct:sign:h':neq:h}, $p_x$ only signs \ta{signature shares} for \ta{$C_m$}, and thus passes the test at \cref{line:condition:sign:hash:SEND}, and reaches \cref{line:v3.1:forward:after:send}, where it disseminates a \forward message. By \cref{subobs:all:correct:send:forward:form}, this message is of the form $\langle \forward, \ta{C_m}, \bull, \{ \sig_s,\sig_x \} \rangle$, and is valid. As above, by definition of the message adversary, this \forward message is received by at least $c-d>0$ correct nodes. By definition these nodes belong to $\frwdReceiveSet$, which yield $|\frwdReceiveSet|>0$ and $\frwdReceiveSet\neq\emptyset$.
	\end{obsProof}
	
	\begin{subobservation}\label{subobs:all:correct:rv:FW:do:send:FW}
		Any $p_u \in \frwdReceiveSet$ broadcasts a 
		$\langle \forward,\allowbreak \ta{C_m},  \bull,$ $\{\sig_s,\sig_u\} \rangle$ message, where $\sig_s $ and $\sig_u$ are $\psender$ and $p_u$'s valid \ta{signature shares} for~\ta{$C_m$}, respectively.
	\end{subobservation}
	
	
	
	\renewcommand{\obscnt}{\ref{subobs:all:correct:rv:FW:do:send:FW}}
	\begin{obsProof}
		Let $p_u\in \frwdReceiveSet$ upon receiving a valid \forward message $\langle \forward,\allowbreak \ta{C_m},  \bull, \sigsset \rangle$ at \cref{line:arrival:FORWARD}. 
		By the condition of \cref{line:v3.1:test:forward:sent}, $p_u$ has either previously sent a \forward message at \cref{line:v3.1:bcast:forward} or it will send such a message at~\cref{line:v3.1:forward:after:forward}. 
		In both cases, \cref{subobs:all:correct:send:forward:form} applies and guarantees that this message contains \ta{$C_m$} and both $p_u$'s and $\psender$'s valid \ta{signature shares}.
	\end{obsProof}
	
	{Note that $\frwdReceiveSet$ is defined over an entire execution of \cref{alg:3.1:merkle:trees}. \cref{subobs:all:correct:rv:FW:do:send:FW} therefore states that any correct node $p_u$ that receives a valid \forward message \emph{at some point of its execution} also broadcasts a matching \forward message \emph{at some point of its execution}. The two events (receiving and sending a \forward message) might, however, occur in any order. For instance, $p_u$ might first receive a \send message from \psender at \cref{ln:sendRx:SEND}, disseminate a \forward message as a result at \cref{line:v3.1:forward:after:send}, and later on possibly receive a \forward message from some other node at \cref{line:arrival:FORWARD}. Alternatively, $p_u$ might first receive a \forward message at \cref{line:arrival:FORWARD}, and disseminate its own \forward message at \cref{line:v3.1:forward:after:forward} as a result. In this second case, $p_u$ might also eventually receive a \send message from \psender (at \cref{ln:sendRx:SEND}). If this happens, $p_u$ will disseminate a second \forward message at \cref{line:v3.1:forward:after:send}. A correct node, however, never disseminates more than two \forward messages (at \cref{line:v3.1:forward:after:send,line:v3.1:forward:after:forward}).}
	
	\begin{subobservation}\label{subobs:all:forwd:hm:received:by:Fdown}
		Any broadcast of 
		$\langle \forward, \ta{C_m}, \frag, \{ \sig_s,\sig_u \} \rangle$
		by a correct $p_u \in \frwdReceiveSet$ arrives to at least $c-\Dadv$ correct nodes that are each, eventually, in $\frwdReceiveSet$.
	\end{subobservation}
	
	
	\renewcommand{\obscnt}{\ref{subobs:all:forwd:hm:received:by:Fdown}}
	\begin{obsProof}
		Each broadcast by a correct $p_u$ of a \forward message
		is eventually received by at least $c-\Dadv$ correct nodes by definition of the message adversary.
		By \cref{subobs:all:correct:send:forward:form} the \forward message contains~\ta{$C_m$}, by \cref{lem:all:msgs:sent:valid} it is valid. Thus, each of its at least $c-\Dadv$ correct recipients belong in $\frwdReceiveSet$, by definition.
		%
		%
		%
	\end{obsProof}
	
	{Because \forward messages are disseminated at \cref{line:v3.1:forward:after:forward}, the reception and sending of \forward messages by correct nodes will induce a ``chain reaction'' until a correct node is reached that has already disseminated a \forward message.
        This ``chain reaction'' mechanism is the intuitive reason why some correct node will eventually receive enough distinct \forward messages to trigger an MBRB-delivery, as captured by the following observation.}
	
	\begin{subobservation}\label{subobs:exist:correct:c-d:sigs}
		There exists a correct node~$p_j$ that receives messages 
		$\langle \forward, \ta{C_m},\bull, \allowbreak\sigsset_u=\allowbreak\{\sig_s,\allowbreak\sig_u\} \rangle$
		from at least $(c-\Dadv)$ distinct correct nodes $p_u$, 
		where $\sig_s=(\ta{\TSIGN_s(C_m)},s) $ and $\sig_u=(\ta{\TSIGN_u(C_m)},u)$ are $\psender$ and $p_u$'s valid \ta{signature shares} for~\ta{$C_m$}, respectively,
		and the \forward message is the \emph{last} message sent by $p_u$.
	\end{subobservation}
	
	\newcommand{\frwdBcastSet}{F_{\downarrow}}
	
	\renewcommand{\obscnt}{\ref{subobs:exist:correct:c-d:sigs}}
	\begin{obsProof}
		By \cref{subobs:all:correct:rv:FW:do:send:FW}, any nodes $p_u\in \frwdReceiveSet$ broadcasts at least one message $\langle \forward, \ta{C_m}, \bull, \sigsset_u=\allowbreak\{\sig_s,\allowbreak\sig_u\} \rangle$, that includes $p_u$'s valid \ta{signature share} for~\ta{$C_m$}, $\sig_u=$ $\big(\ta{\TSIGN_u(C_m)},u\big)$.
		Consider all the \forward messages 
		sent by nodes in $\frwdReceiveSet$ during the \emph{last time} they perform such a broadcast. 
		By \cref{subobs:all:forwd:hm:received:by:Fdown}, there are $|\frwdReceiveSet|$ senders, $p_u \in \frwdReceiveSet$, such that each of $p_u$'s last broadcast of 
		a \forward message
		is guaranteed to be delivered to at least $c-\Dadv$ correct nodes~$p_x$, such that eventually $p_x \in \frwdReceiveSet$.
		Thus, at least $|\frwdReceiveSet|(c-\Dadv)$ such messages are received by nodes in~$\frwdReceiveSet$, overall. 
		By \cref{subobs:frwdReceiveSet:not:empty}, $\frwdReceiveSet$ contains at least one node. We can, therefore, apply the pigeonhole principle, where $\frwdReceiveSet$ are the holes and the above $|\frwdReceiveSet|(c-\Dadv)$ messages are the pigeons, and observe that there exists a node~$p_j\in \frwdReceiveSet$ that will receive at least $|\frwdReceiveSet|(c-\Dadv)/|\frwdReceiveSet|$ such messages. 
		Since we limit the discussion to a \emph{single}, \ie the last, broadcast performed by each node in $\frwdReceiveSet$, 
		no node in~$\frwdReceiveSet$ receives two of the above messages that were originated by the same node in~$\frwdReceiveSet$.
		Therefore, we deduce that $p_j$ has received messages of the form $\langle \forward, \ta{C_m}, \bull, \sigsset_u \rangle$ from at least $(c-\Dadv)$ \emph{distinct} correct nodes $p_u$ and the \forward message is the \emph{last} message sent by $p_u$.
	\end{obsProof}
	
	\begin{subobservation}\label{subobs:one:correct:delivers:m}
		At least one correct node MBRB-delivers $m$ from $\psender$.
	\end{subobservation}
	
	\renewcommand{\obscnt}{\ref{subobs:one:correct:delivers:m}}
	\begin{obsProof}
		By \cref{subobs:exist:correct:c-d:sigs}, there is a correct node~$p_j$ that receives messages of the form $\langle \forward, \ta{C_m}, \bull, \sigsset_u \rangle$ from at least $(c-\Dadv)$ distinct correct nodes $p_u$, such that these \forward messages are the \emph{last} message sent by each $p_u$. Let us denote by $U$ the set of such nodes~$p_u$, hence,  $|U|\ge c-\Dadv$.
		
		Still by \cref{subobs:exist:correct:c-d:sigs}, $p_j$ receives a valid \ta{signature share} $\sig_u=(\ta{\TSIGN_u(C_m)},u)$ from each node $p_u\in U$. It thus receives at least $(c-\Dadv)$ \emph{distinct} \ta{signature shares} for~\ta{$C_m$}.
		\cref{asm:MBRBassumptionR} says $3t + 2{\Dadv} < n$, and thus, $n + 3t + 2{\Dadv} < 2n$ and $n + t < 2n - 2t - 2{\Dadv}$.
		Since $n - t \leq c$, we have $(n+t)/ 2 < n - t - {\Dadv} \leq c - {\Dadv}$. 
		Thus, $p_j$ receives more than $(n+t )/2 $ valid distinct \ta{signature shares}
		for~\ta{$C_m$}.
		
		Let us now consider the set of correct nodes $S$ that receive the initial \send messages disseminated by $\psender$ at \cref{ln:v3.1:commV1ldotsVn}. Any node $p_x\in S$ receives through the \send message its allocated fragment $(\tilde m_x, \pi_x,x)$ from~\psender. By definition of the message adversary, the \send messages disseminated at \cref{ln:v3.1:commV1ldotsVn} are received by at least $c - {\Dadv}$ correct nodes, therefore $|S|\geq c - {\Dadv}$. Furthermore, all nodes in $S$ broadcast a \forward message at \cref{line:v3.1:forward:after:send}, and this will be their \emph{last} \forward message, due to the condition of~\cref{line:v3.1:test:forward:sent}. 
		By the above reasoning, this \forward message will contain their message fragment, that is, it will be of the form $\langle \forward, \ta{C_m}, (\tilde m_x, \pi_x,x), \sigsset_u \rangle$. 
		By \cref{lem:all:msgs:sent:valid}, they are all valid.

		By definition of $S$ and $U$, both these sets  contain only correct nodes, 
		thus, $|S \cup U| \le c$.
		As a result, $|S\cap U|=|S| + |U| - |S\cup U|\geq 2\times (c-\Dadv) -c = c -2\Dadv$. 
		The last \forward messages broadcast by nodes in $S\cap U$ are received by $p_j$ by the definition of $U$. 
		As argued above about nodes in~$S$ (and thus applying to nodes in $S\cap U$), \forward
		messages sent by a node in $S\cap U$  contain their valid message fragment and proof of inclusion~$(\tilde m_x, \pi_x,x)$. 
		It follows that $p_j$ accumulates at least $c -2\Dadv$ distinct such message fragments with their (valid) proof of inclusion.  By \cref{asm:MBRBassumptionR}, $c -2\Dadv\geq \ECCthreshold$.
		
		
		To conclude the proof, note that we have shown that $p_j$ eventually receives more than $(n+t )/2 $ valid distinct \ta{signature shares} for~\ta{$C_m$}, and additionally, that $p_j$ accumulates at least $k$ valid message fragments with their proof of inclusion. At this point, the condition of \cref{line:quorum:sigs:enough:fragments}  becomes \true for~$p_j$.
		Because the \ta{commitment is collision-resistant (\cref{sec:vc}), once $C_m$} is fixed, 
		we can assume that, except with negligible probability, all the message fragments that $p_j$ has received correspond to the fragments computed by \psender at \cref{line:fragments:for:m}.
		By the parameters of the ECC we use, it can recover the message $m$ from any $\ECCthreshold$ or more (correct) fragments generated by~$\psender$, where missing fragments are considered as erasures.
        Therefore, the message $m_j$ reconstructed at \cref{line:v3.1:reconstruct:m} by $p_j$ is the message initially MBRB-broadcast by $\psender$. As a result $m_j=m$, and $p_j$ MBRB-delivers~$m$ at \cref{ln:MBRBdeliverMjSn}.
	\end{obsProof}
	
\end{lemmaProof}

\cref{thm:Global} is the detailed version of \cref{thm:SGlobal}.

\begin{lemma}[MBRB-Global-delivery]
	\label{thm:Global}
	Suppose a correct node, $p_i$, MBRB-delivers an {application message}~$m$.
	{At least $\ell=\displaystyle c - \Dadv\left(\frac{1}{1-\left(\frac{\ECCthreshold-1}{c-\Dadv}\right)}\right)$ correct nodes MBRB-deliver~$m$.}
\end{lemma}
\renewcommand{\lemcnt}{\ref{thm:Global}}
\begin{lemmaProof}
	Suppose a correct node $p_i$ MBRB-delivers $m$ (\cref{ln:MBRBdeliverMjSn}).
    Let us denote by \ta{$C_m$ the commitment returned by $\computeFragCom(m)$}.
	%
		%
		The proof follows a counting argument on the \bundle messages disseminated by correct nodes at \cref{line:v3.1:bundle:comm,line:v3.1:bundle:with:bot}. In the following, if a correct node disseminates a \bundle message both at \cref{line:v3.1:bundle:comm,line:v3.1:bundle:with:bot}, we only consider the one from \cref{line:v3.1:bundle:comm}.

		\begin{subobservation}\label{subobs:all:bundle:valid:hm}
			All valid \bundle messages exchanged during the execution of \cref{alg:3.1:merkle:trees} contain~\ta{$C_m$, the commitment} of the message~$m$, where $m$ is the message MBRB-delivered by~$p_i$.
		\end{subobservation}
		
		\renewcommand{\obscnt}{\ref{subobs:all:bundle:valid:hm}}
		\sloppy\begin{obsProof}	
			Consider a valid message $\langle \bundle,\allowbreak \ta{C'}, \frag'_j,\allowbreak \frag'_i, \ta{\Sigma'} \rangle$.
            By definition of a valid \bundle message, \ta{$\Sigma'$ aggregates a set $\sigsset'$ of $\tau=\lfloor\frac{n+t}{2}\rfloor+1$ valid signature shares for $C'$.}
            \ta{Similarly, when $p_i$ MBRB-delivers $m$ at \cref{ln:MBRBdeliverMjSn}, it has a threshold signature $\Sigma_m$ which aggregates a set $\sigsset_m$ of $\tau=\lfloor\frac{n+t}{2}\rfloor+1$ valid signature shares for $C_m$.}
            By a reasoning identical to that of \cref{thm:duplicity}, these two inequalities imply that $\sigsset'\cap \sigsset_m$ contains the \ta{signature shares} from at least one common correct node, $p_u$.
            As \ta{signature shares} cannot be forged, $p_u$ has issued \ta{signature shares} for both \ta{$C'$ and $C_m$}, and by \cref{lem:single:hash:sign}, \ta{$C'=C_m$}. 
			To complete the proof, note that by the definition of a valid \bundle message, the threshold signature it contains is valid with respect to the \ta{commitment} it carries.
            Hence, all valid \bundle messages must contain the \ta{commitment~$C_m$} of the application message~$m$ that matches the \ta{threshold signature $\Sigma'$} they contain. 
		\end{obsProof}
		
		Let \bundleSendSet be the set of correct nodes that disseminate at least one \bundle message during their execution. Similarly, let \bundleReceiveSet be the set of correct nodes that receive at least one valid \bundle message from a correct node during their execution. The following holds.
		
		

		\medskip
		\cref{subobs:bundleR:and:bundleS:geq:c-d} is a detailed version of~\cref{subobs:SbundleR:and:bundleS:geq:c-d} in \cref{thm:SGlobal}.
		
		\begin{subobservation}\label{subobs:bundleR:and:bundleS:geq:c-d}
			$c \geq |\bundleReceiveSet| \geq c-\Dadv$ and $c \geq |\bundleSendSet| \geq c-\Dadv$.
		\end{subobservation}
		
		\renewcommand{\obscnt}{\ref{subobs:bundleR:and:bundleS:geq:c-d}}
		\begin{obsProof}
			Since \bundleSendSet and \bundleReceiveSet contain only correct nodes, trivially $c \geq |\bundleSendSet|$ and $c \geq |\bundleReceiveSet|$.
			Since $p_i$ MBRB-delivers $m$ at \cref{ln:MBRBdeliverMjSn},  it must have disseminated \bundle messages of the form $\langle \bundle, \ta{C_m}, (\tilde m'_i,\pi'_i,i), (\tilde m'_j,\pi'_j,j), \ta{\Sigma_C} \rangle$ at \cref{line:v3.1:bundle:comm}. 
			The \bundle messages sent by~$p_i$ eventually reach at least $c-\Dadv$ correct nodes, as the message adversary can remove at most~$\Dadv$ of these \bundle messages. 
			By \cref{lem:all:msgs:sent:valid}, these \bundle messages are valid. 
			Hence, $\bundleReceiveSet\ge c-\Dadv>0$ proves the lemma's first part.

			The nodes in $\bundleReceiveSet$ (which are correct) execute \cref{line:arrival:BUNDLE,line:isValid:in:BUNDLE}, and reach \cref{line:test:if:already:bundle}. 
			Because $p_i$ has included a non-$\bot$ second fragment in all its \bundle message, any of the $(c-d)$ nodes of $\bundleReceiveSet$ that receive one of $p_i$'s \bundle messages and has not already sent a \bundle message passes the condition at \cref{line:test:if:already:bundle}.
			%
			%
			Each such node then disseminates a (valid) \bundle message at \cref{line:v3.1:bundle:with:bot}. This behavior
			yields $|\bundleSendSet|\geq c- {\Dadv}$. 
		\end{obsProof}

		Let \bundleKReceiveSet be the set of correct nodes that receive \bundle messages from at least $\ECCthreshold$ distinct correct nodes.

		\medskip
		\cref{subobs:counting:bundle:equation} is a detailed version of~\cref{subobs:Scounting:bundle:equation} in \cref{thm:SGlobal}.
  
		\begin{subobservation}\label{subobs:counting:bundle:equation}
			$|\bundleKReceiveSet|\times|\bundleSendSet| + (\ECCthreshold-1)(|\bundleReceiveSet|-|\bundleKReceiveSet|) \geq |\bundleSendSet|(c-\Dadv)$.
		\end{subobservation}
		
		\begin{figure}[ht]
			\centering
			\begin{tikzpicture}[scale=0.85]
    \tikzmath{
    \heightRcvk = 2.5;
    \heightB = 1.5;
    \widthRcvk = 3;
    \widthRcv = 6;
    }
    
    \def\colorA{black}
    \def\colorB{black}

    \draw[-{>[scale=2.5,length=2,width=3]}] (0,-.25) -- (0,\heightRcvk+.5);
    \node[align=center] at (-2.75,\heightRcvk) {\# received\\\bundle};
    \draw[-{>[scale=2.5,length=2,width=3]}] (-.25,0) -- (\widthRcv+1,0);
    \node[align=center] at (\widthRcv+2,0) {\# correct\\nodes};
    \draw (\widthRcv,0) -- (\widthRcv,-.1);
    
    \node at (-.7,\heightRcvk) {$|\bundleSendSet|$};
    \draw (-.1,\heightRcvk) -- (\widthRcvk,\heightRcvk) -- (\widthRcvk,-.1);
    \node at (\widthRcvk,-.4) {$|\bundleKReceiveSet|$};
    
    \node at (-.7,\heightB) {$\ECCthreshold-1$};
    \draw (0,\heightB) -- (-.1,\heightB);
    \draw[dashed] (0,\heightB) -- (\widthRcvk,\heightB);
    \draw (\widthRcvk,\heightB) -- (\widthRcv,\heightB) -- (\widthRcv,-.1);
    \node at (\widthRcv,-.4) {$|\bundleReceiveSet|$};
\end{tikzpicture}
			\caption{Distribution of distinct \bundle messages received by correct nodes.
				The proof of \cref{subobs:bundleKReceiveSet:geq:frac} shows that $|\bundleSendSet|>\ECCthreshold-1$.}
		\label{fig:msg_dist}
	\end{figure}

	\renewcommand{\obscnt}{\ref{subobs:counting:bundle:equation}}
	\begin{obsProof}
		Let us denote by \numberBundleMsg the overall number of valid \bundle messages received by correct nodes from distinct correct senders. More specifically, in the case when a correct node disseminates \bundle messages both at \cref{line:v3.1:bundle:comm,line:v3.1:bundle:with:bot}, we only consider the \emph{last} \bundle message, \ie the one of \cref{line:v3.1:bundle:comm}. 
		We know that each $p\in \bundleSendSet$ sends a \bundle message, which by \cref{lem:all:msgs:sent:valid} is valid.
            As the message adversary may drop up to~$\Dadv$ out of the $n$ messages of this $\comm$, we are guaranteed that at least $c-\Dadv$ correct nodes receive $p$'s \bundle message. This immediately implies that   
		\begin{equation}\label{eq:numberBundleMsg:geq}
			\numberBundleMsg \geq |\bundleSendSet|(c-\Dadv).
		\end{equation}
		
		
		As illustrated in \cref{fig:msg_dist}, the nodes in $\bundleKReceiveSet$ may receive up to $|\bundleSendSet|$ valid \bundle messages from distinct correct senders (one from each sender in $\bundleSendSet$), for a maximum of $|\bundleKReceiveSet|\times|\bundleSendSet|$ \bundle messages overall. The remaining nodes of $\bundleReceiveSet\setminus\bundleKReceiveSet$ may each receive up to $\ECCthreshold-1$ valid \bundle messages from distinct correct senders, by definition of $\bundleKReceiveSet$. 
		As $\bundleKReceiveSet\subseteq\bundleReceiveSet$ by definition, $|\bundleReceiveSet\setminus\bundleKReceiveSet|=|\bundleReceiveSet|-|\bundleKReceiveSet|$, and the nodes of $\bundleReceiveSet\setminus\bundleKReceiveSet$
		accounts for up to $(\ECCthreshold-1)(|\bundleReceiveSet|-|\bundleKReceiveSet|)$ 
		valid \bundle messages overall. As the \bundle messages counted by \numberBundleMsg are received either by correct nodes in $\bundleKReceiveSet$ or in $\bundleKReceiveSet\setminus\bundleReceiveSet$, these observations lead to
		\begin{equation}\label{eq:numberBundleMsg:leq}
			|\bundleKReceiveSet|\times|\bundleSendSet| + (\ECCthreshold-1)(|\bundleReceiveSet|-|\bundleKReceiveSet|)   \geq \numberBundleMsg.
		\end{equation}
		
		Combining \cref{eq:numberBundleMsg:geq,eq:numberBundleMsg:leq} yields the desired inequality.
	\end{obsProof}

		\medskip
		\cref{subobs:bundleKReceiveSet:geq:frac} is a detailed version of~\cref{subobs:SbundleKReceiveSet:geq:frac} in \cref{thm:SGlobal}.

	\begin{subobservation}\label{subobs:bundleKReceiveSet:geq:frac}
		$\displaystyle |\bundleKReceiveSet| \geq c - \Dadv\left(\frac{1}{1-\left(\frac{\ECCthreshold-1}{c-\Dadv}\right)}\right).$
	\end{subobservation}

	\renewcommand{\obscnt}{\ref{subobs:bundleKReceiveSet:geq:frac}}
	\begin{obsProof}
		Rearranging the terms of~\cref{subobs:counting:bundle:equation}, and recalling that $|\bundleReceiveSet|\leq c$ and $\ECCthreshold\ge 1$, we get
		\begin{equation}
			|\bundleKReceiveSet|\times(|\bundleSendSet|-\ECCthreshold+1) \geq |\bundleSendSet|(c-\Dadv) - |\bundleReceiveSet|(\ECCthreshold-1)  \geq |\bundleSendSet|(c-\Dadv) -c(\ECCthreshold-1).
		\end{equation}
		By \cref{subobs:bundleR:and:bundleS:geq:c-d} and \cref{asm:MBRBassumptionR}, $|\bundleSendSet| \geq c-\Dadv \geq c-2 \Dadv\geq \ECCthreshold$, therefore $|\bundleSendSet|-\ECCthreshold+1>0$, and the previous equation can be transformed in
		\begin{equation}\label{eq:bundleKReceiveSet:geq}
			|\bundleKReceiveSet| \geq
			\frac{|\bundleSendSet|(c-\Dadv)-c(\ECCthreshold-1)}{|\bundleSendSet|-\ECCthreshold+1}.
		\end{equation}
		Note that the right-hand side of \cref{eq:bundleKReceiveSet:geq} is a monotone increasing function in~$|\bundleSendSet|$ when $|\bundleSendSet|>\ECCthreshold-1$, as its derivative, 
		$\frac{\Dadv(\ECCthreshold+1)}{(|\bundleSendSet|-\ECCthreshold+1)^2}$, is positive.
		By \cref{subobs:bundleR:and:bundleS:geq:c-d}, $|\bundleSendSet|\in [c-\Dadv, c]\subseteq[\ECCthreshold,c]$. 
		The minimum of the right-hand side of \cref{eq:bundleKReceiveSet:geq} is therefore obtained for $\bundleSendSet=c-\Dadv$, yielding
		\begin{equation}
			|\bundleKReceiveSet| \geq
			\frac{(c-\Dadv)^2-c(\ECCthreshold-1)}{(c-\Dadv)-\ECCthreshold+1}
			= c - \Dadv\left(\frac{1}{1-\left(\frac{\ECCthreshold-1}{c-\Dadv}\right)}\right).
		\end{equation}
	\end{obsProof}

	\medskip
	\cref{subobs:all:bundleKReceiveSet:mbrb:deliv:m} is a detailed version of~\cref{subobs:Sall:bundleKReceiveSet:mbrb:deliv:m} in \cref{thm:SGlobal}.

	\begin{subobservation}\label{subobs:all:bundleKReceiveSet:mbrb:deliv:m}
		All nodes in $\bundleKReceiveSet$ MBRB-deliver~$m$.
	\end{subobservation}
	
	\renewcommand{\obscnt}{\ref{subobs:all:bundleKReceiveSet:mbrb:deliv:m}}
	\begin{obsProof}
		Consider $p_u\in \bundleKReceiveSet$. By the definition of $\bundleKReceiveSet$, the node~$p_u$ receives $\ECCthreshold$ valid \bundle messages from $\ECCthreshold$ distinct correct nodes. Let us denote by $\langle \bundle,\allowbreak \ta{C_x}, (\tilde m_x,\pi_x,x), \bull, \ta{\Sigma_x} \rangle $ these $\ECCthreshold$ messages with $x\in[\ECCthreshold]$. By \cref{subobs:all:bundle:valid:hm}, for all $x\in[\ECCthreshold]$, \ta{$C_x=C_m$}.
        In addition, \ta{$p_u$ stores each received threshold signature $\Sigma_x$, which is valid for $C_m$}.
		
		Because the messages are valid, so are the proofs of inclusions $\pi_x$, and as we have assumed that the \ta{commitments are collision-resistant, $C_x=C_m$} implies that the received fragments $\tilde m_x$ all belong to the set of fragments computed by \psender at \cref{line:fragments:for:m} for $m$. As the \bundle messages were received from $\ECCthreshold$ distinct correct nodes, the node~$p_u$ receives at least $\ECCthreshold$ distinct valid fragments for $m$ during its execution.
		%
		%
		If $p_u$ has not MBRB-delivered any message yet, the condition at \cref{line:quorum:sigs:enough:fragments} eventually becomes true for \ta{$C_m$}, and $p_u$ reconstructs $m$ at \cref{line:v3.1:reconstruct:m}, since it possesses at least $k$ (correct) message fragments, which are sufficient for the correct recovery of~$m$ by our choice of ECC.
        Then, $p_u$ MBRB-delivers $m$ at \cref{ln:MBRBdeliverMjSn}.
		On the other hand, if $p_u$ has already MBRB-delivered some message~$m'$,
		then \cref{thm:duplicity} (MBRB-No-duplicity) implies $m'=m$, since 
		$p_i$ is known to have MBRB-delivered~$m$. 
		Therefore, in all possible cases, $p_u$ MBRB-delivers~$m$.
	\end{obsProof}
	
	\cref{thm:Global} follows from \cref{subobs:bundleKReceiveSet:geq:frac,subobs:all:bundleKReceiveSet:mbrb:deliv:m} and the fact that all nodes in $\bundleKReceiveSet$ are correct.
\end{lemmaProof}	


\Subsection{Discussion: Selection of~\texorpdfstring{$\ECCthreshold$}{k}}
\label{sec:select-k}
In the above analysis, we set $\ECCthreshold$ to be a parameter that controls the number of fragments that allow decoding the $\ECC$. To obtain the communication depicted in \cref{sec:MBRBscheme}, we assumed $\ECCthreshold=\Omega$($n$). Furthermore, this parameter affects the delivery power of the MBRB algorithm, as seen in \cref{thm:Global}, namely $\ell=\displaystyle c - \Dadv\left({1-\left(\frac{\ECCthreshold-1}{c-\Dadv}\right)}\right)^{-1}$.

Let us assume that we wish to design an MBRB algorithm with a specified delivery power of $\ell=c-(1+\eps)\Dadv$, for some $\eps>0$.
Plugging in \cref{thm:Global}, we need the delivery power~$\ell$ provided by the \cref{alg:3.1:merkle:trees} to surpass $c-(1+\eps)\Dadv$, thus
\begin{align*}
	c-(1+\eps)\Dadv \le  c - \Dadv\left(\frac{1}{1-\left(\frac{\ECCthreshold-1}{c-\Dadv}\right)}\right)
\end{align*}
leading to 
\(
\ECCthreshold\le \frac{\eps}{1+\eps}(c-\Dadv)+1\text{.}
\) 
That is, choosing any integer
\(
\ECCthreshold \le \frac{\eps}{1+\eps}(n-t-\Dadv) +1 
\), 
satisfies the above.
Recall that the blowup of the ECC is given by $\nECC/\kECC \approx n/k$ (\cref{sec:ECC-instantiation}), which implies that for any application message~$m$, we have $|\ECC(m)|\approx \frac{n}{k}|m| = \frac{1+\eps}{\eps}\cdot\frac{n}{n-t-\Dadv}|m|$.

Together with \cref{asm:MBRBassumptionR}, we conclude that the constraints on~$k$ that support delivery power of $\ell \ge n-t-(1+\eps)\Dadv$, are 
\[
k \le \min \left\{
n-t-2\Dadv,
\frac{\eps}{1+\eps}(n-t-\Dadv) +1
\right\}\text{.}
\]

\Subsection{Supporting multiple instances and multiple senders}
\label{sec:multi}
We remark that the above analysis fits the single-shot broadcast algorithm with a fixed sender.
As mentioned above, a multi-shot multi-sender algorithm can be achieved by communicating the identity of the sender and a sequence number along with any piece of information communicated or processed during this algorithm.
This added information uniquely identifies any piece of information with the respective instance.
Additionally, \ta{signature shares, threshold signatures, commitments, and proofs of inclusion} should be performed on the application message~$m$ augmented with the sender's identifier and the sequence number.
This will prevent Byzantine nodes from using valid \ta{signature shares/threshold signatures} from one instance in a different instance. 
As a result, an additive factor of $\bigO{\log n}$ bits has to be added to each communicated message, which yields additive communication of $\bigO{n^2\log n}$ and has no effect on the asymptotic communication, as we explained in the proof of \cref{lem:communication}.
Other changes, such as augmenting the application message $m$ with the sender's identifier and sequence number do not affect the length of \ta{signature shares, threshold signatures, commitments, and proofs of inclusion}.

\section{Using Bracha's BRB on hash values under a message adversary}
\label{sec:Bracha:with:MA}

\ft{%
Das, Xiang, and Ren~\cite{DBLP:conf/ccs/DasX021} have proposed a communication optimal BRB algorithm that does not use signatures and relies on Bracha instead to reliably broadcast a hash value of the initial sender's message.
One might legitimately ask whether this approach could not be easily adapted to withstand a message adversary, possibly resulting in an MBRB algorithm exhibiting optimal communication complexity (up to the size of hashes $\kappa$), optimal Byzantine resilience ($n>3t+2d$), and optimal delivery power $n-t-d$ (or at least some close-to-optimal delivery power $\ell$, up to some factor $\epsilon$).

\newcommand{\lbrb}{\ensuremath{\ell_{\mathit{BRB}}}\xspace}
\newcommand{\ECHO}{\textsc{echo}\xspace}
\newcommand{\READY}{\textsc{ready}\xspace}

Unfortunately, under a message adversary, Bracha's BRB leads to a sub-optimal Byzantine resilience, and degraded delivery power $\ell$. In particular, Albouy, Frey, Raynal and Taiani~\cite{DBLP:conf/opodis/AlbouyFRT22} have shown that Bracha can be used to implement a MBRB algorithm, but their solution requires a sub-optimal resilience bound ($n > 3t + 2d + 2\sqrt{td}$) and yields a reduced delivery power $\lbrb= \left\lceil n-t-\big(\frac{n-t}{n-3t-d}\big)d \right\rceil$.
Disappointingly, these less-than-optimal properties would in turn be passed on to any MBRB algorithm using Bracha's BRB along the lines of Das, Xiang, and Ren's solution.\footnote{Taking into account the initial dissemination of the message $m$ by the broadcaster, which is also impacted by the message adversary, such an algorithm could in fact at most reach a delivery power of $\max\big(0,(n-t-d)+\lbrb-(n-t)\big)=\max\big(0,\lbrb-d\big)=\max\big(0,\left\lceil n-t-\big(\frac{n-t}{n-3t-d}+1\big)d \right\rceil\big)$.}
By contrast, the algorithm we propose is optimal in terms of communication cost (up to $\kappa$) and Byzantine resilience, and close to optimal in terms of delivery power (up to some parameter $\epsilon$ that can be chosen arbitrarily small).

To provide a hint of why Bracha's BRB leads to degraded resilience and delivery power when confronted with a message adversary (MA), consider the classical \ECHO phase of Bracha's BRB~\cite{B87}. At least one correct node must receive $(n+t)/2$ \ECHO messages to ensure the first \READY message by a correct node can be emitted. To ensure Local-delivery, the threshold $(n+t)/2$ must remain lower than the worst number of \ECHO messages a correct node can expect to receive when the sender is correct. Without an MA this constraint leads to $(n+t)/2<n-t$, which is verified by assuming $n>3t$. With an MA, the analysis is more complex. Applying a similar argument to that of the proof of \cref{thm:SGlobal}, one can show that in the worst case the adversary can ensure that no correct node receives more than $(n-t-d)^2/(n-t)$ \ECHO messages. Ensuring that at least one correct node reaches the Byzantine quorum threshold $(n+t)/2$ requires therefore that
\begin{equation*}
     (n+t)/2< (n-t-d)^2/(n-t).
\end{equation*}

This leads to a quadratic inequality involving $n$, $t$ and $d$, which results in the following constraint on $n$:
\begin{equation*}
n > 2t + 2d + \sqrt{(d+t)^2+d^2} \geq 3t + 3d.
\end{equation*}

In their analysis~\cite{DBLP:conf/opodis/AlbouyFRT22}, Albouy, Frey, Raynal, and Taiani improve on this resilience bound by systematically optimizing the various retransmission and phase thresholds used by Bracha's BRB algorithm, but their solution still falls short of the optimal lower bound $n > 3t + 2d$, which the solution presented in this paper provides. 
}

\end{document}